\documentclass{article}
\title{Two types of branching programs with bounded repetition that cannot efficiently compute monotone 3-{\sc cnf}s}
\author{Igor Razgon \\Department of Computer Science and Information Systems\\ 
Birkbeck University of London \\igor@dcs.bbk.ac.uk}
\usepackage{amssymb}
\usepackage{amsmath}
\usepackage{comment}
\usepackage{graphicx}
\newtheorem{observation}{Observation}
\newtheorem{claim}{Claim}
\newtheorem{myexample}{Example}
\newtheorem{theorem}{Theorem}
\newtheorem{lemma}{Lemma}
\newtheorem{definition}{Definition}
\date{}
\begin{document}
\maketitle
\begin{abstract}
It is known that there are classes of 2-CNFs requiring exponential size non-deterministic
read-once branching programs to compute them. However, to the best of our knowledge,
there are no superpolynomial lower bounds for branching programs of a higher repetition
computing a class of 2-CNFs. This work is an attempt to make a progress in this direction.

We consider a class of monotone 3-CNFs that are almost 2-CNFs in the sense that in each clause there
is a literal occurring in this clause only. We prove exponential lower bounds for two classes
of non-deterministic branching programs.
The first class significantly generalizes monotone read-$k$-times {\sc nbp}s and the second
class generalizes oblivious read $k$ times branching programs.
The lower bounds remain exponential for $k \leq \log n/a$
where $a$ is a sufficiently large constant. 

\begin{comment}
We prove exponential lower bounds for two classes of non-deterministic branching programs ({\sc nbp}s)
with bounded repetition computing a class of functions representable by monotone 3-{\sc cnf}s.
The first class significantly generalizes monotone read-$k$-times {\sc nbp}s and the second
class generalizes oblivious read $k$ times branching programs. To the best of our knowledge,
this is the first separation of monotone {\sc nbp}s with bounded repetition from
monotone 3-{\sc cnf}s as well as the first separation of oblivious read-$k$-times branching
programs (even determinisitc ones) from 3-{\sc cnf}s. The lower bounds remain exponential for $k \leq \log n/a$
where $a$ is a sufficiently large constant. 
\end{comment}
\end{abstract}
\section{Introduction}
{\bf Statement of Results.}
\begin{comment}
Although exponential lower bounds for non-deterministic branching programs ({\sc nbp})s 
with bounded repetition or bounded path length are well known (e.g. \cite{...}), 
to the best of our knolwedge, such ({\sc nbp})s have
not been separated yet from ({\sc cnf})s. In particular, we believe there is no yet a result 
showing a super-polynomial lower bound for (syntactic) read $k$-times {\sc nbp}s computing a class
of functions representable as {\sc cnf}s with polynomial nuber of clauses.
\end{comment}
%%%%%%%%%%%%%%%%%%%%%%%%%%%%%%%%%%%%%%%%%%New Introduction%%%%%%%%%%%%%%%%%%%%%%%%%%%%%%%%%%%%%%%%%%%%%%%%%%%%%%%%%%%%%
It is known that there are classes of 2-CNFs that require Nondeterministic Read-Once Branching Programs (NROBPs) 
of exponential size for their representation. 
For example, in \cite{RazgonIPEC14}, we have essentially shown that a NROBP
computing a monotone 2-CNF with a bounded number of occurrences of each variable is of size exponential
in the pathwidth of the primal graph of the 2-CNF. An exponential lower bound is thus easy to obtain by
taking a 2-CNF corresponding to an expander graph.

A natural direction for further research is to understand the complexity of branching programs with higher 
repetition on monotone 2-CNFs. Indeed, to the best of our knowledge, for 2-CNFs, there are no superpolynomial 
lower bounds for non-deterministic branching programs with a repetition higher than one, even for the restricted
cases where the branching programs are allowed to be monotone or oblivious and deterministic. 

This work is an attempt to make a progress in this direction. Instead of monotone 2-CNFs we consider \emph{padded}
monotone 3-CNFs that are `almost' 2-CNFs. 
In particular each clause of a padded monotone 3-CNF contains a \emph{padding} literal, that is a literal appearing in this clause
only, the number of occurrences of the rest two literals is larger than $1$.
It is not hard to see that there is a one-to-one correspondence between padded monotone 3-CNFs and graphs
without isolated vertices or vertices of degree $1$. In particular, the CNF corresponding to such a graph $G$
has $V(G) \cup E(G)$ as the set of variables and the clauses correspond to the edges. In particular, 
literals of the clause corresponding to $e \in E(G)$ are the variable corresponding to $e$ and the variables
corresponding to the end vertices of $e$. We denote the CNF corresponding to $G$ by $CNF(G)$.

In this paper we consider the class of CNFs $CNF(K_n)$ where $K_n$ is the complete graph of $n$ vertices.
We estalish exponential lower bounds for non-deterministic read-$k$-times monotone branching programs and non-determinisitc read-$k$-times
oblivous branching programs where $k \leq \log m/c$ where $m$ is the number of variables of the considered
CNF and $c$ is a sufficiently large constant.

In fact, in order to `push' as far as possible towards exponential lower bounds for non-deterministic read-$k$-times
branching programs, we establish the exponential lower bounds for generalizations of non-deterministic monotone and
oblivious read $k$-times branching programs that are specified below.

Let us parition the variables of $CNF(G)$ into \emph{vertex variables}, i.e. those that correspond ot the vertices
of $G$ and \emph{edge variables} )those that correspond to the edges of $G$.
The generalization of non-deterministic read-$k$-times monotone branching programs is called 
$k$-Vertex Edge Monotone Branching Programs ($k$-{\sc vembp}s).
In this model vertex variables are allowed to appear at most $k$ times (monotonicity is \emph{not} required),
while edge variables are required to be monotone, though the number of their occurrences is not limited.

The generalization of non-deterministic oblivious read-$k$ times branching programs
requires vertex variables to be oblivious and to occur at most $k$ on each \emph{consistent
path} (that is, this restriction is \emph{semantic}, while for $k$-{\sc vembp}, the restriction
is \emph{syntactic} because it applies to each path). The edge variables are not constrained at all.
We refer to this model $k$ vertex oblivious branching programs ($k$-{\sc vobp}).

{\bf Overview of the proof.}
As mentioned above,
the considered class of {\sc cnf}s is associated with (undirected simple) graphs 
as follows. Given a graph $G$, the {\sc cnf} $CNF(G)$ \cite{RazgonKR} has variables $\{X_u|u \in V(G)\}$
(the vertex variables) and
$\{X_{u,v}|\{u,v\} \in E(G)\}$ (the edges variables) and the clauses $(X_u \vee X_{u,v} \vee X_v\}$ for each
$\{u,v\} \in E(G)$. The above class ${\bf K}$ consists of $CNF(K_n)$ for even $n$ where
$K_n$ is a complete graph of $n$ vertices.

The lower bounds are stated in Theorem \ref{mainlower}.
In order to prove the theorem, we introduce a new graph parameter, which we call
$(k,c)$-fold matching width. This is a generalization of matching width which we used to
prove lower bounds in \cite{RazgonKR,RazgonIPEC14,RazgonAlgo}, in particular, the matching width can be seen as $(1,2)$-fold
matching width. Putting $c=8k+10$ and using Theorem 1.1. from \cite{AMaass}, we show 
(Theorem \ref{cliquebound}) that the $(k,c)$-fold matching width of 
clique is at least $n/2^c$. Then we show (resp. Theorems \ref{mainupgraded} and \ref{mainoblivious}) that 
for a graph $G$ with $(k,c)$-fold matching width
at least $t$, both $k$-{\sc vembp} and $k$-{\sc vobp} need an exponential size in 
$t/(c-1)$ to compute $CNF(G)$. Theorem \ref{mainlower} follows from combination
of Theorems \ref{cliquebound}, \ref{mainupgraded}, and \ref{mainoblivious}.

In order to prove Theorem \ref{mainupgraded}, we associate a matching $M$ of $G$ with
an (essentially) disjunction of {\sc cnf} which we call a decision tree w.r.t. $M$.
The main part of the proof is Theorem \ref{manytrees} stating that if ${\bf A}$
is a set of decision trees w.r.t. matching of size $t$ such that every satisfying assignment of
$CNF(G)$ satisfies at least one element of ${\bf A}$ then $|{\bf A}|$ must be exponential in $t$.
We prove this by a probabilistic argument, defining a probability space over satisfying assignments
of $CNF(G)$ and showing that the probability of satisfying an individual decision tree w.r.t. a matching
of size at least $t$ is exponentially small in $t$. Then Theorem \ref{manytrees} immediately follows
by the union bound. Theorem \ref{mainupgraded} is then proved by showing that it is possible to define
a function from a subset of length $c-1$ sequences of nodes of a $k$-{\sc vembp} $Z$ computing $CNF(G)$ to decision trees
w.r.t matchings of size at least $t$ so that every satisfying assignment of $CNF(G)$ would satisfy a decision
tree in the range of this function. It will immediately follow from Theorem \ref{manytrees} that size of the range
is exponential in $t$ and hence so is the size of the domain (upper-bounded by the number of $c-1$-sequences of nodes of $Z$) 
of the function. We will conclude that the number of nodes of $Z$ is exponential in $t/(c-1)$.

To prove Theorem \ref{mainoblivious}, we specify a set ${\bf S}$ of $2^t$ satisfying assignments of $CNF(G)$.
For each $S \in {\bf S}$, we specify a consistent root-leaf path $P^S$ of a $k$-{\sc vobp} $Z$ conputing
$CNF(G)$ whose labelling set of literals is a subset of $S$. Then, using `superposition' of paths,
a standard `fooling' technique for proving lower bounds for oblivious branching programs, we identify
on each $P^S$ at most $c-1$ nodes such that the respective vectors of these nodes do not coincide for
any $P^{S_1}$ and $P^{S_2}$ with $S_1 \neq S_2$. Then we apply the counting argument similar to the one
mentioned in the end of the previous paragraph.

{\bf Related work.}
Exponential lower bounds for branching programs with bounded repetition are well known
both when occurrences of each individual variable are restricted \cite{readktimes}
and the total length of root-leaf paths is restricted \cite{Ajtai99}. 

The author considered monotone padded $3$-CNFs in \cite{cobdd} for 
non-deterministic semantic $k$-OBDD, a special case on nondeterministic semantic oblivious read $k$-times
branching programs where the sequence of variables on each
computational path is a subsequence of $k$ copies of a \emph{fixed} permutation.
We are not aware of other results separating branching programs of bounded repetition
from monotone $2$-CNFs or monotone padded $3$-CNFs.

The only other results we are aware of that separate branching programs with repetition higher than $1$
and CNFs with constant length clauses are \cite{Jukna4clique} (for nondeterministic read-$k$-times branching
programs with $k \leq c\log /\log \log n$ for some constant $c$) and \cite{Sauerhoff03}
(for randomized read-$k$-times branching programs with repetition $k \leq c\log n$ for some constant $c$ 
and probability of error exponentially small in $2^k$).
Both these results consider CNFs with all literals being negative, so the respective
lower bounds clearly hold for monotone CNFs resulting from `switching sign' of literals.

Definition of monotone {\sc nbp}s can be found in \cite{MonComp} or \cite{RazFCT}, we are not aware
of previous results considering monotone {\sc nbp}s with bounded repetition.

\section{Preliminaries}
In this paper by a \emph{set of literals} we mean one that does not
contain both an occurrence of a variable and its negation.
For a set $S$ of literals we denote by $Var(S)$ the set of variables
whose literals occur in $S$ (the $Var$ 
notation naturally generalizes to {\sc cnf}s and Boolean functions).
A set $S$ of literals
represents the truth assignment to $Var(S)$ where variables occurring
positively in $S$ (i.e. whose literals in $S$ are positive) are assigned with $true$
and the variables occurring negatively are assigned with $false$.
For example, the assignment $\{x_1 \leftarrow true, x_2  \leftarrow true, x_3 \leftarrow false\}$
to variables $x_1,x_2,x_3$ is represented as $\{x_1,x_2,\neg x_3\}$.

Let $CC$ be a {\sc cnf}. A set $S$ of literals \emph{satisfies} a clause $C$ of $CC$
if at least one literal of $C$ belongs to $S$. If all clauses of $CC$ are satisfies by $S$ then
$S$ \emph{satisfies} $CC$. If, in addition, $Var(CC)=Var(S)$ then we say that $S$ is a 
\emph{satisfying assignment} of $CC$. The notion of a satisfying assignment naturally extends
to Boolean functions $F$ meaning a truth assignment to $Var(F)$ on which $F$ is true.

We are now going to present a general terminology related to Non-deterministic branching
programs ({\sc nbp}s). An {\sc nbp} $Z$ computing a Boolean function $F$ is a directed acyclic 
graph ({\sc dag}) with one root and one leaf with some edges labelled with literals of 
variables of $Var(F)$. Multiple edges are allowed. 
A (directed) path $P$ of $Z$ is \emph{consistent} if opposite literals
of the same variable do not occur on it (as labels of its edges). For a consistent path $P$,
we denote by $A(P)$ the set of literals labelling its edges. A consistent root-leaf path of
$Z$ is called a \emph{computational path} of $Z$. The connection between $Z$ and $F$ is the
following. A set $S$ of literals with $Var(S)=Var(F)$ is a satisfying assignment of $F$
if and only if there is a computational path $P$ of $Z$ such that $A(P) \subseteq S$.
The \emph{size} of $Z$ denoted by $|Z|$ is the number of its vertices. %nodes??
A well known way to obtain special classes of {\sc nbp}s is by imposing constraints on its
root-leaf paths (e.g. the number occurrences of each variable). If such a restriction is
imposed on \emph{computational} paths only, this restriction is called \emph{semantic}.
Otherwise, if the restriction is imposed on \emph{all} root-leaf paths, it is called 
\emph{syntactic}.

{\bf Remark.} 
Note that if $Z$ computes a {\sc cnf} $CC$ then, for every computational path $P$,
$A(P)$ satisfies $CC$. Indeed, otherwise, $A(P)$ can be extended to an assignment 
to all the variables of $Var(CC)$ that falsifies one of its clauses.

\section{The lower bounds}
%The class of {\sc cnf}s we consider for the purpose of obtaining lower bounds
%correspond to (simple undirected) graphs. 
Let $G$ be a graph.
Then by $CNF(G)$ we denote the monotone 3-{\sc cnf} having the set of variables 
$VVar(G) \cup EVar(G)$ where $VVar(G)=\{X_u|u \in V(G)\}$ and
$EVar(G)=\{X_{u,v}| \{u,v\} \in E(G)\}$ and the set of clauses
$\{(X_u \vee X_{u,v} \vee X_v)|\{u,v\} \in E(G)\}$. We call variables of $VVar(G)$
and $EVar(G)$ the \emph{vertex variables} and the \emph{edge variables} of $CNF(G)$,
respectively.

Now we define two types of {\sc nbp}s computing $CNF(G)$.
The first type has the following constraints: (i) on \emph{each} root-leaf path, 
each vertex variable occurs at most $k$ times; (ii) all the occurrences of edge variables 
on \emph{each} root-leaf path are positive (the branching program is monotone on edge variables)
but their repetition is unbounded.
\begin{comment}
\begin{itemize}
\item On \emph{each} root-leaf path, each vertex variable occurs at most $k$ times.
\item All the occurrences of edge variables on \emph{each} root-leaf path are positive (the branching program is monotone on edge variables)
      but their repetition is unbounded. 
\end{itemize}
\end{comment}
Note that the above constraints are \emph{syntactic} because they apply to each root-leaf path of 
of the considered branching program.
We call a branching program obeying the above constraints a $k$-\emph{Vertex Edge Monotone Branching Program} 
($k$-{\sc vembp}). Note that $k$-{\sc vembp}s generalize monotone read-$k$-times {\sc nbp}s.

The second considered class of {\sc nbp}s obeys the following constraints:
(i) on each \emph{computational} path, each vertex variable occurs at most $k$ times;
(ii) there is a sequence $XSV$ of $VVar(G)$  where each variable occurs exactly $k$ times 
such that the sequence of vertex variables along each computational path is a subsequence of $XSV$.
\begin{comment}
\begin{itemize}
\item On each \emph{computational} path, each vertex variable occurs at most $k$ times.
\item There is a sequence $XSV$ of $VVar(G)$  where each variable occurs exactly $k$ times 
such that the sequence of vertex variables along each computational path is a subsequence of $XSV$.
\end{itemize}
\end{comment}
Note that the above constraints are \emph{semantic} because they apply to computational paths
only. Moreover, there are no constraints on edge variables \emph{even} on computational paths.
We call an {\sc nbp} obeying the above constraints a $k$-Vertex-Oblivious Branching Program
($k$-{\sc vobp}). Note that a $k$-{\sc vobp} generalizes semantic oblivious read-$k$-times 
{\sc nbp}s.

Now we are ready to state the lower bounds for the above branching programs.
For that purpose, recall that $K_n$ is a clique on $n$ variables.

\begin{theorem} \label{mainlower}
Let $s=8k+10$.
Then the size of $k$-{\sc vembp} computing $CNF(K_n)$ is at least
$(8/7)^{\Omega(n/(2^s*(s-1)))}$ and the size of $k$-{\sc vobp} computing
is at least $2^{n/(2^s*(s-1))}$. %the factor 1/n$ is due to the uniforming
\end{theorem}

Note that the number $m$ of variables of $CNF(K_n)$ is $n(n+1)/2$.
That is, for sufficiently large $n$, $m^{1/2}<n<2m^{1/2}$ and hence,
in terms of the number of variables, the above lower bounds
can be restated as $(8/7)^{\Omega(m^{1/2}/(2^s*(s-1)))}$ and $2^{m^{1/2}/(2^s*(s-1))}$,
respectively, for a sufficiently large $n$. 
Clearly the lower bound remains exponential for $k \leq \log n/a$ where
$a$ is a sufficiently large constant. 
%Insert something about logarithmic bounds on $k$ for the above lower bound
%to remain exponential

In order to prove Theorem \ref{mainlower}, we introduce a graph parameter
and show this parameter is large for cliques and that both 
$k$-{\sc vembp} and $k$-{\sc vobp} are of exponential size on $CNF(G)$
when this parameter for $G$ is large. Theorem \ref{mainlower}
will immediately follow from the combination of these two statements.
The parameter is a generalization of matching width that we used in
\cite{RazgonIPEC14} to prove a lower bound for read-once branching programs.

An \emph{interval} of a sequence $S=s_1, \dots, s_q$ is a subsequence $s_i, \dots s_j$
($1 \leq i \leq j \leq q$). 
\begin{definition} 
{\bf $c$-separation.}
Let $S$ be a sequence of elements of a universe set $U$. 
Let $S_1, \dots, S_c$ be a partition of $S$ into intervals.
Suppose $X,Y \subseteq U$ are disjoint sets such that, for each $1 \leq i \leq c$,
the following is true: (i) $S_i \cap (X \cap Y) \neq emptyset$;
(ii) if $i$ is odd then $S_i \cap (X \cap Y) \subseteq X$;
(iii) if $i$ is even then $S_i \cap (X \cap Y) \subseteq Y$. 
\begin{comment}
\begin{itemize}
\item $S_i \cap (X \cap Y) \neq emptyset$
\item If $i$ is odd then $S_i \cap (X \cap Y) \subseteq X$.
\item If $i$ is even then $S_i \cap (X \cap Y) \subseteq Y$. 
\end{itemize}
\end{comment}
Then $S$ has a $c$-\emph{separation} w.r.t. $(X,Y)$ and $S_1, \dots, S_c$
are \emph{witnessing intervals} of this separation. 
\end{definition}

\begin{myexample}
Let $U=\{1,2,3,4,5\}$ and let $S=(1,2,3,1,4,5,1,1,3,2)$.
Then the intervals $S_1=(1,2,3),S_2=(1,4,5,1),S_3=(1,3,2)$
witness a $3$-separation of $S$ w.r.t. $(\{2,3\},\{4,5\})$. 
\end{myexample}

\begin{definition} \label{manyfold}
A graph $G$ has $c_1,c_2$-\emph{fold} matching width at least $t$
if for each sequence $SV$ of elements of $V(G)$ where each element appears exactly 
$c_1$ times, there is a matching $M=\{\{u_1,v_1\}, \dots, \{u_t,v_t\}\}$ of $G$ such
that $SV$ has a $c \leq c_2$-separation w.r.t. $(U=\{u_1, \dots, u_t\},
V=\{v_1, \dots, v_t\})$. 

An equivalent but, possibly, more intuitive definition is that for each
sequence $SV$ of elements of $V(G)$ where each element occurs at most $c_1$
times there is a partition of $SV$ into at most $c_2$ intervals and
a matching $M$ as above so that there is no interval containing elements of both $U$ and $V$. 
\end{definition}

\begin{theorem} \label{cliquebound}
Let $s=8k+10$. Then for each even $n$, the $(k,s)$-fold matching width of $K_n$
is at least $\lfloor n/2^s \rfloor$.
\end{theorem}

{\bf Proof.}
Let $SV$ be a sequence of vertices of $K_n$ where each vertex appears exactly 
$k$ times. Let $U$, $V$ be two disjoint subsets of $V(G)$. An interval $I$
of $SV$ is a \emph{link} between $U$ and $V$ if intermediate elements of 
$I$ do not belong to $U \cup V$ and either
(i) the first element of $I$ belong to $U$ and the last element of $I$ belongs to $V$
or (ii)  the first element of $I$ belong to $V$ and the last element of $I$ belongs to $U$.

\begin{claim}
There are disjoint sets $U$, $V$ of size at least  $\lfloor n/2^s \rfloor$
such that there are at most $c \leq s-1$ links between them. 
\end{claim}

{\bf Proof.}
Assume the claim is not true. Put $\ell=\lfloor n/2^s \rfloor$.
Let $\{1, \dots, n\}$ be elements of $V(G)$ being arbitrarily enumerated and
let $U_1=\{1, \dots, n/2\}$ and $U_2=\{n/2+1, \dots, n\}$.
By our assumption, for each $U \subseteq U_1$, $V \subseteq U_2$ 
with $|U|=|V|=\ell$ there are at least $s$ links between $U$ and $V$.
By Theorem 1.1. of Alon and Maass \cite{AMaass}, 
$|SV| \geq n(s-9)/8=n(8k+10-9)/8=nk+n/8>nk$ in contradiction to the definition of $SV$.
$\square$

Let $U=\{u_1, \dots, u_{\ell}\}$ and let $V=\{v_1, \dots, v_{\ell}\}$ 
be as stated in the claim and let $I_1, \dots, I_c$ be the
links between them in $SV$. Note that $M=\{\{u_1,v_1\}, \dots, \{u_t,v_t\}\}$ is a matching of $K_n$.
It remains to show that there is $c+1$ separation either w.r.t.$(V,U)$ or w.r.t. $(U,V)$.
Let $x_1, \dots, x_c$ be the respective first
elements of $I_1, \dots, I_c$. It is not hard to see that they are all distinct elements of $SV$.
Assume w.l.o.g. that they occur on $SV$ in the order they are listed.
Let $SV_1, \dots, SV_{c+1}$ be partitioning of $SV$ into intervals defined as follows.
$SV_1$ is the prefix of $SV$ whose last element is $x_1$. For $1<i \leq c$, $SV_i$
is an interval whose first element is the immediate successor of $x_{i-1}$ and
the last element is $x_c$. Finally, $SV_{c+1}$ is the suffix of $SV$ starting at the
immediate successor of $x_c$. Observe that $SV_1, \dots, SV_{c+1}$ is well defined.
In particular, notice that $SV_{c+1}$ is not empty because it contains the last
element of $I_c$. 

Notice that by construction, each $SV_i$ contains an end vertex of a link between $U$ and $V$ and hence
intersects with either $U$ or $V$. On the other, no $SV_i$ intersects with \emph{both}
$U$ and $V$ because no $SV_i$ contains a whole $U-V$-link. It follows that 
each $SV_i$ is either `$U$-only' or `$V$-only. Next, by construction,
any $SV_j+SV_{j+1}$ contains $I_j$ and hence intersects with both $U$ and $V$. 
It follows that if $SV_j$ is $U$-only then $SV_{j+1}$  is $V$-only and vice versa.
We conclude that either all $SV_j$ with odd $j$ are $U$ only and all $SV_j$ with even $j$ are $V$-only
or, all $SV_j$ with odd $j$ are $V$-only and all $SV_j$ with even $j$ are $U$ only.
It follows that $SV_1, \dots, SV_{c+1}$ are witnessing intervals of $c+1 \leq s$ separation of $SV$ 
w.r.t. either $(U,V)$ or $(V,U)$. $\blacksquare$

\begin{theorem} \label{mainupgraded}
Let $G$ be a graph of $n$ vertices.
Let $k,c,t$ be integers such that the $k,c$-fold matching of $G$ is at least $t$. 
Let $Z$ be a $k$-{\sc vembp} implementing $CNF(G)$. Then $Z$ has at least $(8/7)^{\Omega(t/(c-1))}$ nodes. 
\end{theorem}

\begin{theorem} \label{mainoblivious}
Let $G,n,k,c,t$ be as in Theorem \ref{mainupgraded} and let 
$Z$ be a $k$-{\sc vobp} implementing $CNF(G)$. 
Then $Z$ has at least $2^{t/(c-1)}$ nodes. 
\end{theorem}

Theorems \ref{mainupgraded} and \ref{mainoblivious} are proved in Sections \ref{freesec} and
\ref{oblsec}, respectively. Now, we already to prove Theorem \ref{mainlower}.

{\bf Proof of Theorem \ref{mainlower}.}
Put $c=8k+10$ and $t=n/2^c$.
Then the hypotheses of both Theorem \ref{mainupgraded} and \ref{mainoblivious}
are satisfied by Theorem \ref{cliquebound}.
Theorem \ref{mainlower} now immediately follows from Theorems 
\ref{mainupgraded} and \ref{mainoblivious} by substitution $t=n/2^c$.
$\blacksquare$

\section{Proof of Theorem \ref{mainupgraded}} \label{freesec}
A $k$-{\sc vembp} is \emph{uniform} if on each root-leaf path
each \emph{vertex} variable occurs \emph{exactly} $k$ times.
The following lemma is proved in Section A of the Appendix. 

\begin{lemma} \label{unigood}
Let $Z$ be a $k$-{\sc vembp} computing $CNF(G)$ for a graph $G$.
Then there is a uniform $k$-{\sc vembp} of size $O(n^2|Z|^2k)$ computing
$CNF(G)$, where $m$ is the number of edges of $G$.
\end{lemma}

We prove Theorem \ref{mainupgraded} for a uniform $k$-{\sc vembp}
as it follows from Lemma \ref{unigood} that this assumption does
not restrict generality. 
%Indeed, what is an upper bound on $m$
%in terms of $n$ and $Z$? The number of multiple edges between any
%two particular vertices $u$ and $v$ of $Z$ can be assumed $O(n)$
%because if there are two multiple edges unlabelled or two labelled
%with the same literal then one of them can be safely removed. 
%Then $m=O(nZ^2)$. It follows that for any $k$-{\sc vembp}
%there is a uniform $k$-{\

%Below, we, in fact prove
%the $(8/7)^{t/(c-1)}$ lower bound for a uniform $k$-{\sc vembp}
%and the result of Theorem \ref{mainupgraded} immediately follows
%by introducing the $1/n$ correcting factor.The following is a useful
%fact about monotone $k$-{\sc vembp}s.

\begin{lemma} \label{samevargen}
Let $Z$ be a uniform $k$-{\sc vembp} implementing $CNF(G)$.
Let $P_1$ and $P_2$ be two paths having the same initial and final vertices.
Then $Var(P_1) \cap VVar(G)=Var(P_2) \cap VVar(G)$. Put it differently,
a vertex variable $X_u$ occurs on $P_1$ if and only if $X_u$ occurs on $P_2$. 
\end{lemma}

{\bf Proof.}
Let $u$ and $v$ be the starting and ending vertices of $P_1$ and $P_2$.
Denote by $rt$ and $lf$ the root and leaf vertices of $Z$.
Let $P_0$ be a $rt-u$ path of $Z$ and $P_3$ be a $v-lf$ path of $Z$.
Then, due to the acyclicity both $Q_1=P_0+P_1+P_3$ (the concatenation of $P_0,P_1,P_3$)
and $Q_2=P_0+P_2+P_3$ are root-leaf paths of $Z$. 
Suppose that $V_1=Var(P_1) \cap VVar(G)  \neq V_2=Var(P_2) \cap VVar(G)$
and that, say, $V_1 \setminus V_2 \neq \emptyset$.
Let $X_u \in V_1 \setminus V_2$. Since $X_u$ does not occur in $P_2$, it occurs
$k$ times in $P_1 \cup P_3$ due to the uniformity of $Z$, in particular, requiring
$k$ occurrences of $X_u$ on $Q_2$. It follows that in $Q_1$ $X_u$ occurs
at least $k+1$ times in contradiction to the definition of a $k$-{\sc vembp}.
$\blacksquare$

\begin{definition} \label{mcnf}
{\bf ${\bf MCNF(M)}$ and $XIR(M)$.}
Let $G$ be a graph and let $M=\{\{u_1,v_1\}, \dots, \{u_t,v_t\}\}$ be a matching of $G$.
A \emph{matching} {\sc cnf} w.r.t. $M$ consists
of $t$ clauses $C_1, \dots C_t$ such that $C_i$ is either 
$(X_{u_i} \vee X_{u_i,v_i})$  or $(X_{u_i,v_i} \vee X_{v_i})$. 
We denote the set of all matching {\sc cnf}s w.r.t. $M$ by ${\bf MCNF}(M)$.
We denote the set $\{X_u|u \in V(G) \setminus V(M)\}$ by $XIR(M)$
(`IR' stands for `irrelevant'). 
\end{definition}

\begin{definition}\label{destree}
{\bf Decision tree w.r.t. $M$.}
Let $G$ and $M$ be as in Definition \ref{mcnf}.
A decision tree $T$ w.r.t. $M$ is a directed tree whose edges 
are oriented from the root towards the leaf. 
Each leaf node $a$ is associated with an element of ${\bf MCNF}(M)$
denoted by $\varphi(a)$. Each non-leaf node $a$ is associated with
a variable of $XIR(M)$ denoted by $Var(a)$ so that on any root-leaf
path each variable of $XIR(M)$ occurs at most once as a label of
a non-leaf node. Each non-leaf node $a$ has two outgoing edges called
the \emph{positive} and \emph{negative} edges that are associated with the respective
literals of $Var(a)$ (i.e. $Var(a)$ and $\neg Var(a)$, respectively).
For a root-leaf path $P$ of $T$ we denote by $L(P)$ the set of literals
associated with the edges of $P$. A set $S$ of literals \emph{satisfies} $T$
if there is a root-leaf path $P$ of $T$ such that $L(P) \subseteq S$
and $S$ satisfies $\varphi(a)$ where $a$ is the last vertex of $P$.
\end{definition}

\begin{figure}[h]
\includegraphics[height=5cm]{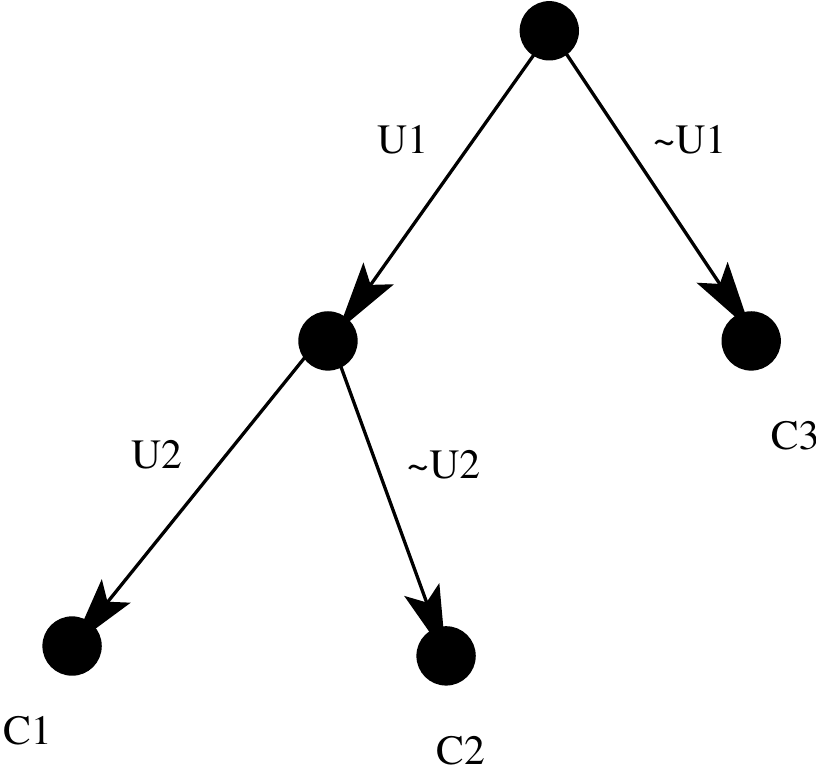}
\caption{Illustration of a desision tree}
\label{destreepic}
\end{figure}

\begin{myexample}
Let $G$ be a graph on vertices $\{u_1, \dots, u_6\}$ and suppose that
$M=\{\{u_3,u_4\},\{u_5,u_6\}\}$ is a matching of $G$.
Then $C_1=(X_{u_3} \vee X_{u_3,u_4}) \wedge (X_{u_5} \vee X_{u_5,u_6})$,
     $C_2=(X_{u_3} \vee X_{u_3,u_4}) \wedge (X_{u_6} \vee X_{u_5,u_6})$, and
     $C_3=(X_{u_4} \vee X_{u_3,u_4}) \wedge (X_{u_6} \vee X_{u_5,u_6})$
are all matching {\sc cnf}s w.r.t. $M$.
Consider a decision tree $T$ shown in Figure \ref{destreepic}.
Labels $u_i,\neg u_i$ denote, respectively the positive and negative edges with
a tail $a$ such that $Var(a)=X_{u_i}$. A leaf $b$ labelled with $C_i$ means that
$\varphi(b)=C_i$. 
Observe that the set $S_1=\{X_{u_1}, \neg X_{u_2},X_{u_3},X_{u_6}\}$ satisfies
$T$ because it contains the literals labelling the root-leaf path to $C_2$ and also satisfies
$C_2$. On the other hand, the set $S_2=\{X_{u_1}, X_{u_2}, \neg X_{u_3}, \neg X_{u_3,u_4}\}$
does not satisfy $T$ because the only possible root-leaf path whose literals are contained
in $S_2$ is the one leading to $C_1$; however $S_2$ falsifies $C_1$. 
\end{myexample}

For a directed path $P$, we say that $P_1, \dots, P_c$ is a \emph{partition} of 
$P$ into \emph{subpaths} if $P_1, \dots, P_c$ are subpaths of $P$ such that $P_1$
is a prefix of $P$, $P_c$ is a suffix of $P$ and for any $P_i,P_{i+1}$, the last vertex
of $P_i$ is the first vertex of $P_{i+1}$. If $x_1, \dots, x_{c-1}$ are last vertices of
$P_1, \dots, P_{c-1}$, respectively, then we say that $P_1, \dots, P_c$ is a partition
of $P$ into subpaths w.r.t. $x_1, \dots, x_{c-1}$. 

For a path $P$ of a $k$-{\sc vembp} $Z$ computing $CNF(G)$, 
we denote by $SV(P)$ the sequence of vertices of $G$
listed in the order the respective vertex variables occur along $P$. 
For example if the sequence of literals occurring on $P$
is $(X_{u_1},X_{u_2},X_{u_2},X_{u_3,u_4},X_{u_2},X_{u_1,u_2},X_{u_3})$
then $SV(P)=(u_1,u_2,u_2,u_2,u_3)$ (simply remove the edge variables and
replace each $X_{u_i}$ of the resulting sequence by $u_i$).
The next two lemmas are proved in the following two respective subsections of this section.
\begin{lemma} \label{destreecover}
Let $Z$ be a uniform $k$-{\sc vembp} computing $CNF(G)$
and let $P$ be a computational path of $Z$ (recall that a computational path is a consistent root-leaf path).
Let $M=\{\{u_1,v_1\}, \dots, \{u_t,v_t\}\}$ be a matching of $G$.
Suppose that there is a partition $P_1,\dots, P_c$ of $P$ into subpaths such that
$SV(P_1), \dots, SV(P_c)$ are intervals witnessing a $c$-separation
of $SV(P)$ w.r.t. $(U=\{u_1, \dots, u_t\},V=\{v_1, \dots, v_t\})$. 
Let $x_1, \dots, x_{c-1}$ be the respective end vertices of
$P_1, \dots, P_{c-1}$. Then there is a decision tree $D$ w.r.t. $M$ 
such that for any computational path $Q$ passing through 
$x_1, \dots, x_{c-1}$, $A(Q)$ satisfies $D$. 
\end{lemma}

\begin{lemma} \label{manytrees}
Let ${\bf A}$ be a family of decsion trees w.r.t. matchings of size at least $t$ 
such that any satisfying assignment $S$ of $CNF(G)$ satisfies at least one 
element of ${\bf A}$. Then $|{\bf A}| \geq (8/7)^t$.
\end{lemma}

{\bf Proof of Theorem \ref{mainupgraded}}
Let $P$ be a computational path of $Z$.
Let $(x_1, \dots x_{c'})$ be a sequence of distinct intermediate vertices
of $P$, occurring on $P$ in that order and let $P_1, \dots, P_{c'+1}$
be the partition of $P$ into subpath w.r.t. $x_1, \dots, x_{c'}$.
Suppose there are $U=\{u_1, \dots, u_t\}$, $V=\{v_1, \dots, v_t\}$
and a matching $M=\{\{u_1,v_1\}, \dots, \{u_t,v_t\}\}$ of $G$ such that
$SV(P_1), \dots, SV(P_{c'+1})$ are intervals witnessing separation of $SV(P)$
w.r.t. $(U,V)$. Then we say that $(x_1, \dots x_{c'})$ is a $t$-\emph{separation
vector} of $P$. 

Let ${\bf XV}$ be the set of all sequences of vertices of $Z$ of length at most $c-1$ 
such that each $XV \in {\bf XV}$ is a $t$-separation vector of some computational path
of $Z$. By Lemma \ref{destreecover}, for each $XV \in {\bf XV}$, we can associate a decision
tree $DT(XV)$ w.r.t. a matching of size $t$ such that for any computational path $Q$ of
$Z$ passing through $XV$, $A(Q)$ satisfies $DT(XV)$.

Let ${\bf A}$ be the set of all $DT(XV)$. Observe that each satisfying assignment $S$ of $CNF(G)$
satisfies an element of ${\bf A}$. Indeed, let $P$ be a computational path of $Z$ such that
$A(P) \subseteq S$. By definition of $k,c$-fold matching width, there is a matching
$M=\{\{u_1,v_1\}, \dots, \{u_t,v_t\}\}$ such that there is a $c' \leq c$-separation of $SV(P)$
w.r.t. $(U=\{u_1, \dots, u_t\}, V=\{v_1, \dots, v_t\})$.
Let $SV_1, \dots, SV_{c'}$ be a partition of $SV(P)$ into intervals witnessing the $c$-separation.
It is not hard to see that there is a partition $P_1, \dots, P_{c'}$ of $P$ into subpaths such
that $SV(P_i)=SV_i$ for $1 \leq i \leq c'$. It follows that the sequence $XV=(x_1, \dots, x_{c'-1})$
of respective end vertices of $P_1, \dots, P_{c'-1}$ is a $t$-separation vector of $P$.
Due to the bound on the length, $XV \in {\bf XV}$ and hence, by the previous paragraph,
$A(P)$ satisfies $DT(XV)$ and hence $S$ satisfies $DT(XV)$ as well.
It follows from Lemma \ref{manytrees} that $|{\bf A}| \geq (8/7)^t$.

On the other hand, $|{\bf A}| \leq |{\bf XV}|$ and $|{\bf XV}|$ is at most as the number of sequences
of (not necessarily distinct) vertices of $Z$ of length $c-1$.
That is, $|{\bf A}| \leq |{\bf XV}| \leq |Z|^{c-1}$. Combining this with the previous paragraph,
we obtain $|Z|^{c-1} \geq (8/7)^t$ and the theorem follows. $\blacksquare$
 
\subsection{Proof of Lemma \ref{destreecover}}
Let $Q^1$ and $Q^2$ be two computational paths of $Z$ passing through $x_1, \dots, x_{c-1}$.
Let $Q^1_1, \dots, Q^1_c$ and $Q^2_1, \dots, Q^2_c$ be respective partitions of $Q^1$
and $Q^2$ into subpaths w.r.t. $x_1, \dots, x_{c-1}$. 
Let $Q^*$ be the root-leaf path passing through $x_1, \dots, x_{c-1}$
whose respective partition $Q^*_1, \dots, Q^*_{c}$ into subpaths w.r.t. $x_1, \dots, x_{c-1}$
is as follows. For each $1 \leq i \leq c$, $Q^*_i=Q^1_i$ whenever $i$ is odd and $Q^*_i=Q^2_i$
whenever $i$ is even. We denote $Q^*$ by $Mix(Q_1,Q_2)$.

Using `superposition' of two paths such as $Mix(Q_1,Q_2)$ is a standard `fooling' technique
in proving lower bounds for branching programs. In case of {\sc vembp} such
a technique is not applicable directly because $Mix(Q_1,Q_2)$ may be labelled by opposite literals
of the same variable and thus is not necessarily a \emph{computational} path. However, as shown
in the next lemma, such an approach is possible if one more condition is imposed on $Q_1$ and $Q_2$.

\begin{lemma} \label{computpath}
Suppose $XIR(M) \subseteq Var(A(Q_1) \cap A(Q_2))$ (that is, all variables of $XIR(M)$
are assigned by the same values by both $Q_1$ and $Q_2$). Then $Q^*=Mix(Q_1,Q_2)$ is a
computational path. 
\end{lemma}

{\bf Proof.}
Assume, by contradiction that there is a variable $X$ of $CNF(G)$ such that both
$X$ and $\neg X$ occur as labels of $Q^*$. Then $X$ is not an edge variable because,
due to their monotonicity, edge variables do not occur negatively on $Z$ and also
$X \notin XIR(M)$ by assumption of the lemma. It remains to assume that
$X \in XU \cup XV$ where $XU=\{X_u|u \in U\}$ and $XV=\{X_v|v \in V\}$.
Assume that $X \in XU$, that is $X=X_u$ for some $u \in U$. 
Then the opposite occurrences happen one of $Q^*_i$ and another
on $Q^*_j$ such that $i$ is odd and $j$ is even. Indeed, if $i$ and $j$ are of the same
parity then they both subpaths of either $Q^1$ or $Q^2$ in contradiction to being
$Q^1$ and $Q^2$ computational paths.  It follows that 
$X_u$ occurs on $Q^2_j$. By Lemma \ref{samevargen}, $X_u$ occurs on $P_j$ and hence
$u$ occurs on $SV(P_j)$. However, this is a contradiction to our assumption
that $SV(P_1), \dots, SV(P_c)$ witness a $c$ separation of $SV(P)$ w.r.t. $(U,V)$
(recall that elements of $U$ cannot occur on the `even' intervals of the witness).
The reasoning for the case $X \in XV$ is symmetric.
$\blacksquare$

Using Lemma \ref{computpath}, we can prove a restricted version of Lemma \ref{destreecover}
that then will be used for the induction basis.

\begin{lemma} \label{indbasemat}
Let $S$ be a set of literals of variables of $XIR(M)$.
Then there is a matching {\sc cnf} $CC$ w.r.t. $M$ such that
any computation path $Q$ passing through $x_1, \dots, x_{c-1}$
with $S \subseteq A(Q)$ satisfies $CC$. 
\end{lemma}

{\bf Proof.}
We are going to show that for each each $\{u_i,v_i\}$ of
$M$ there is a clause $C(u_i,v_i) \in \{(X_{u_i} \vee X_{u_i,v_i}),
(X_{u_i,v_i} \vee X_{v_i})\}$ such that for each path $Q$ as in the statement
of the lemma $A(Q)$ satisfies $C(u_i,v_i)$. This will immediately imply
that each $A(Q)$ satisfies the matching {\sc cnf} consisting of clauses
$C(u_1,v_1) \dots C(u_t,v_t)$.

Assume by contradiction that there are computational paths $Q^1$ and $Q^2$
such that $Q^1$ does not satisfy $(X_{u_i} \vee X_{u_i,v_i})$ and
$Q^2$ does not satisfy $(X_{u_i,v_i} \vee X_{v_i})$. Let $Q^*=Mix(Q^1,Q^2)$.
By Lemma \ref{computpath}, $Q^*$ is a computational path and hence $A(Q^*)$
satisfies $(X_{u_i} \vee X_{u_i,v_i} \vee X_{v_i})$ by definition of $Z$.
We derive a contradiction by showing that none of these three literals occurs
in $A(Q^*)$. Indeed, by definition of $Q^1$ and $Q^2$, $X_{u_i,v_i} \notin 
A(Q_1)$ and $X_{u_i,v_i} \notin A(Q_2)$ and hence, clearly, $X_{u_i,v_i} \notin A(Q^*)$.
By the same reasoning as in the proof of Lemma \ref{computpath}, a literal of
$X_{u_i}$ can only occur in $Q^*_i$ for an odd $i$. However, such a $Q^*_i$
is a subpath of $Q^1$ where $X_{u_i}$ does not occur positively. The reasoning regarding
$X_{v_i}$ is symmetric. $\blacksquare$

{\bf Proof of Lemma \ref{destreecover}}
We prove the following more general statement.
Let $S$ be a set of literals with $Var(S) \subseteq XIR(M)$.
Let ${\bf Q}(S)$ be the set of computational paths of $Z$ going 
through $x_1, \dots, x_{c-1}$ such that $S \subseteq A(Q)$.
Then there is a decision tree $DT$ w.r.t. $M$ where 
$Var(S)$ do not occur as labels and such that for each
$Q \in {\bf Q}(S)$, $A(Q)$ satisfies $DT$. The lemma will follow
as a special case with $S=\emptyset$.

The proof is by induction on $|XIR(M) \setminus Var(S)|$.
Assume that $|XIR(M) \setminus Var(S)|=0$, that is $Var(S)=XIR(M)$.
Let $CC$ be a matching {\sc cnf} w.r.t. $M$ satisfied by
$A(Q)$ for all $Q \in {\bf Q}(S)$ according to Lemma \ref{indbasemat}.
Then all these $A(Q)$ satisfy a decision tree with a single node
$a$ such that $\varphi(a)=CC$. 

Assume now that $Var(S) \subset XIR(M)$. If there is $S' \supset S$
with $Var(S')=XIR(M)$ such that ${\bf Q}(S) \subseteq {\bf Q}(S')$
then the previous paragraph applies. Otherwise, there is a variable
$X \in XIR(M) \setminus Var(S)$ such that
${\bf Q}(S)={\bf Q}(S_1) \cup {\bf Q}(S_2)$ where
$S_1=S \cup \{X\}$ and $S_2=S \cup \{\neg X\}$ (recall that, due to
the uniformity, vertex variables and, in particular, variables of
$XIR(M)$, occur on all computational paths of $Z$).

By the induction assumption, there are decision trees $DT_1$
and $DT_2$ that are, respectively related to ${\bf Q}(S_1)$ and
${\bf Q}(S_2)$ as specified in the first paragraph of this proof.
Let $rt_1,rt_2$ be the respective roots of $DT_1$ and $DT_2$.
Let $DT$ be a decision tree obtained from $DT_1$ and $DT_2$ by
introduction of a new vertex $rt$ with $Var(rt)=X$ as the root of $DT$ with $rt_1,rt_2$
being the children of $rt$ and $(rt,rt_1),(rt,rt_2)$ being, respectively,
positive and negative edges of $rt$. It remains to observe that 
$DT$ is satisfied by $A(Q)$ for all $Q \in {\bf Q}(S)$. Indeed, assume 
that $Q \in {\bf Q}(S_1)$ and let $Q'$ be the path of $DT_1$ witnessing that $A(Q)$ satisfies
$DT_1$ (that is $L(Q') \subseteq A(Q)$ and $A(Q)$ satisfies $\varphi(a)$).
Let $Q''$ be the path of $DT$ obtained by appending $Q'$ to $(rt,rt_1)$.
As $X \in S_1 \subseteq A(Q)$, $L(Q'') \subseteq A(Q)$ and hence $Q''$ witnesses
that $A(Q)$ satisfies $DT$. If $Q \in {\bf Q}(S_2)$ then the reasoning is symmetric.
$\blacksquare$

\subsection{Proof of Lemma \ref{manytrees}}
We define a probability space over the set $SAT(G)$ of satisfying assignments of $CNF(G)$
and show that the probability that a decision tree w.r.t. a matching of size at least
$t$ is satisfied by an element of $SAT(G)$ is exponentially small in $t$.
It will follow then that the number of decision trees w.r.t. such matchings must be large
to ensure that each element of $SAT(G)$ satisfies one of them.

We first define a probability space over all possible assignments of $Var(G)$ and then observe that
the assignments having non-zero probabilities are precisely those that belong to $SAT(G)$ 
and hence this space is, in fact, over $SAT(G)$.
A random assignment $S$ in this space is chosen by the following procedure.
For each vertex variable $X$, choose either $X$ or $\neg X$ with probability $1/2$.
Then for each edge variable $X_{u,v}$ choose $X_{u,v}$ with probability $1$ if 
both $\neg X_u$ and $\neg X_v$ have been chosen. Otherwise, choose either $X_{u,v}$ or
$\neg X_{u,v}$ with probability $1/2$. The probability of the chosen assignment is the product 
of probabilities of assignments of individual variables. Due to the way we choose assignments
to edge variables it is not hard to see that indeed, $Pr(S)>0$ if and only if $S \in SAT(G)$.
In particular (although, this is not relevant for the further reasoning),
for each $S \in SAT(G)$, $Pr(S)=(1/2)^{m-ne(S)}$, where $m$ is the number of variables of $CNF(G)$
and $ne(S)$ is the number of edges $\{u,v\}$ such that $\{\neg X_u,\neg X_v\} \subseteq S$. 

An event in the probability space we have just defined is a subset of $SAT(G)$.
We say that a subset $VR$ of variables of $CNF(G)$ is a \emph{prefix} set of variables
if either $VR \subseteq VVar(G)$ or $VVar(G) \subseteq VR$. 
For example, if $G=K_4$ with the set of vertices $\{u_1,u_2,u_3,u_4\}$
then the sets $\{X_{u_3},X_{u_4}\}$ and $\{X_{u_1}, \dots, X_{u_4},X_{u_2,u_3}\}$
are both prefix sets, while the set $\{X_{u_2},X_{u_3},X_{u_1,u_2}\}$ is not a prefix set. 
Let $S$ be a set of literals over a prefix set $VR$ of variables.
We denote by ${\bf EC}(S)$ the event containing all the assignments $S'$ such that $S \subseteq S'$. 
We first need to prove two basic facts related to the ${\bf EC}$ events. The proofs of these facts
are provided in Section B of the Appendix. 

\begin{observation} \label{elem1}
Let $S$ be a set of literals such that $Var(S)$ is a prefix set.
Let $X \notin Var(S)$ be a variable such that $Var(S) \cup \{X\}$
is still a prefix set. 
Let $S_1=S \cup \{X\}$ and $S_2=S \cup \{\neg X\}$. 
Suppose that $X=X_{u,v}$ and both $X_u$ and $X_v$ occur negatively in $S$.
Then $Pr({\bf EC}(S_1)|{\bf EC}(S))=1$ and $Pr({\bf EC}(S_2)|{\bf EC}(S))=0$.
Otherwise, both $Pr({\bf EC}(S_1)|{\bf EC}(S))=0.5$ and
$Pr({\bf EC}(S_2)|{\bf EC}(S))=0.5$.
\begin{comment}
Then the following statements hold.
\begin{itemize}
\item Suppose that $X=X_{u,v}$ and both $X_u$ and $X_v$ occur negatively in $S$.
Then $Pr({\bf EC}(S_1)|{\bf EC}(S))=1$ and $Pr({\bf EC}(S_2)|{\bf EC}(S))=0$.
\item Otherwise, both $Pr({\bf EC}(S_1)|{\bf EC}(S))=0.5$ and
$Pr({\bf EC}(S_2)|{\bf EC}(S))=0.5$.
\end{itemize}
\end{comment}
\end{observation}

\begin{lemma} \label{elem2}
Let $X,S,S_1,S_2$ be as in Observation \ref{elem1}. 
Let ${\bf X}$ be an arbitrary event of ${\bf SAT}(G)$.
Suppose that $X=X_{u,v}$ and both $X_u$ and $X_v$ occur negatively in $S$.
Then $Pr({\bf X}|{\bf EC}(S))=Pr({\bf X}|{\bf EC}(S_1))$.
Otherwise, $Pr({\bf X}|{\bf EC}(S))=Pr({\bf X}|{\bf EC}(S_1))*0.5+Pr({\bf X}|{\bf EC}(S_2))*0.5$.
\begin{comment}
\begin{itemize}
\item Suppose that $X=X_{u,v}$ and both $X_u$ and $X_v$ occur negatively in $S$.
Then $Pr({\bf X}|{\bf EC}(S))=Pr({\bf X}|{\bf EC}(S_1))$.
\item Otherwise, $Pr({\bf X}|{\bf EC}(S))=Pr({\bf X}|{\bf EC}(S_1))*0.5+Pr({\bf X}|{\bf EC}(S_1))*0.5$.
\end{itemize}
\end{comment}
\end{lemma}

\begin{comment}
{\bf Proof.}
It is not hard to see that ${\bf EC}(S)$ is the disjoint union of ${\bf EC}(S_1)$ and
${\bf EC}(S_2)$.
Therefore, by the law of full probability,
$Pr({\bf X} \cup {\bf EC}(S))=Pr({\bf X} \cap {\bf EC}(S_1))+Pr({\bf X} \cap {\bf EC}(S_2))$.
Or, in terms of conditional probabilities
\begin{equation} \label{eq1}
 Pr({\bf X}|{\bf EC}(S))*Pr({\bf EC}(S))=
 Pr({\bf X}|{\bf EC}(S_1))*Pr({\bf EC}(S_1))+Pr({\bf X}|{\bf EC}(S_2))*Pr({\bf EC}(S_2))
\end{equation}

Notice that since ${\bf EC}(S_1) \subseteq {\bf EC}(S)$, we can write
\begin{equation} \label{eq2}
Pr({\bf EC}(S_1))=Pr({\bf EC}(S_1) \cap {\bf EC}(S))=Pr({\bf EC}(S_1)|{\bf EC}(S))*Pr({\bf EC}(S))
\end{equation}

Likewise,
\begin{equation} \label{eq3}
Pr({\bf EC}(S_2))=Pr({\bf EC}(S_2)|{\bf EC}(S))*Pr({\bf EC}(S))
\end{equation}

Substituting (\eqref{eq2}) and (\eqref{eq3}) into the right hand side of (\eqref{eq1})
and dividing both sides by $Pr({\bf EC}(S))$ gives us

\begin{equation} \label{eq4}
Pr({\bf X}|{\bf EC}(S))=
Pr({\bf X}|{\bf EC}(S_1))*Pr({\bf EC}(S_1)|{\bf EC}(S))+Pr({\bf X}|{\bf EC}(S_2))*Pr({\bf EC}(S_2)|{\bf EC}(S))
\end{equation}

The Lemma now follows from Observation \ref{elem1} by choosing the appropriate $Pr({\bf EC}(S_1)|{\bf EC}(S))$
and $Pr({\bf EC}(S_2)|{\bf EC}(S))$ for each considered case.
$\blacksquare$
\end{comment}

In the rest of the proof, we, essentially, first show that the probability of satisfying a matching
{\sc cnf} is exponentially small and then, using induction, extend this statement to decision trees.
However, for the induction to take off, we need to define a generalization of a matching {\sc cnf},
which we call a \emph{matching {\sc cnf} with a tail}. This is a {\sc cnf} $CC$ whose set of clauses
can be partitioned into two subsets $MT(CC)$ and $TL(CC)$, respectively referred to as
the \emph{matching clauses} and the \emph{tail clauses}. $MT(CC)$ is a matching {\sc cnf} w.r.t. a matching $M$
and the clauses of $TL(CC)$ are singletons each contains a literal of a variable of $XIR(M)$. 
For such a $CC$, we denote by ${\bf ES}(CC)$ the event consisting of all the elements of $SAT(G)$ satisfying $CC$.
We are going to bound from above the probability of ${\bf ES}(CC)|{\bf EC}(S)$ where $S$ is an assignment to a prefix set 
of variables. Let us first extend the notation.

A clause $C$ of $CC$ is \emph{satisfied} by $S$
if $S$ contains a literal of $C$ and \emph{falsified} if $S$ contains negations of all the literals of $S$. If $C$ is neither
satisfied nor falsified, it is \emph{retained} by $S$. A retained clause $(X_u \vee X_{u,v})$ of $MT(CC)$ is \emph{resolved}
w.r.t. $S$ if $\neg X_v \in S$. We denote by $DET(CC,S)$ the set consisting of all retained tail clauses and retained non resolved
matching clauses of $CC$ w.r.t. $S$. 

Let $C=(X_u \vee X{u,v})$ be a retained unresolved clause of $CC$ w.r.t. $S$. If $\neg X_u \notin S$,
then $C$ is a \emph{binary clause}  w.r.t. $S$, otherwise, it is a \emph{half clause} w.r.t. $S$. 
If $X_v \in S$ then $(X_u \vee X_{u,v})$ is a
\emph{free clause} w.r.t. $S$, otherwise, it is \emph{constrained clause} w.r.t. $S$. 
We denote by $R_{tl}(CC,S),R_{bc}(CC,S),R_{bf}(CC,S),R_{hc}(CC,S),R_{hf}(CC,S)$
the respective sets of retained tail, binary constrained, binary free,
half constrained and half free clauses of $CC$ w.r.t. $S$. Clearly, the non-empty sets $R_j(CC,S)$, $j \in \{tl,bc,bf,hc,hf\}$
form a partition of $DET(CC,S)$.

\begin{myexample}
Let $G$ be a graph on vertices $u_1, \dots, u_{15}$
and suppose that\\ $M=\{\{u_4,u_5\},\{u_6,u_7\},\{u_8,u_9\},\{u_{10},u_{11}\}\}$.
Consider the following {\sc cnf}.
\begin{equation}
CC=(X_{u_1}) \wedge (\neg X_{u_2}) \wedge (\neg X_{u_3}) \wedge
   (X_{u_4} \vee X_{u_4,u_5}) \wedge (X_{u_6} \vee X_{u_6,u_7}) \wedge
   (X_{u_9} \vee X_{u_8,u_9}) \wedge (X_{u_{11}} \vee X_{u_{10},u_{11}})
\end{equation}
Clearly, it is a matching {\sc cnf} with a tail.
In particular, $MT(CC)=(X_{u_4} \vee X_{u_4,u_5}) \wedge (X_{u_6} \vee X_{u_6,u_7}) \wedge
   (X_{u_9} \vee X_{u_8,u_9}) \wedge (X_{u_{11}} \vee X_{u_{10},u_{11}})$
is a matching {\sc cnf} w.r.t. $M$ and $TL(CC)=(X_{u_1}) \wedge (\neg X_{u_2}) \wedge (\neg X_{u_3})$.
Let $S=\{X_{u_1},X_{u_2},\neg X_{u_5},\neg X_{u_6},X_{u_7},X_{u_8}\}$.
Then clauses $(X_{u_1})$ and $(\neg X_{u_2})$ are, respectively, satisfied and falsified,
clause $(X_{u_4} \vee X_{u_4,u_5})$ is resolved and 
$DET(CC,S)=\{(\neg X_{u_3}), (X_{u_6} \vee X_{u_6,u_7}),
   (X_{u_9} \vee X_{u_8,u_9}), (X_{u_{11}} \vee X_{u_{10},u_{11}})\}$.
Finally, $R_{tl}(CC,S)=\{(\neg X_{u_3})\}$, $R_{hf}(CC,S)=\{(X_{u_6} \vee X_{u_6,u_7})\},
R_{bf}(CC,S)=\{(X_{u_9} \vee X_{u_8,u_9})\}, R_{bc}(CC,S)=\{(X_{u_{11}} \vee X_{u_{10},u_{11}})\}$,
and $R_{hc}(CC,S)=\emptyset$
\end{myexample}
We define $weight(CC,S)$ as follows.
\begin{equation}
weight(CC,S)=(7/8)^{|R_{bc}(CC,S)|}*(3/4)^{|R_{bf}(CC,S)|+|R_{hc}(CC,S)|}*(1/2)^{|R_{hf}(CC,S)|+|R_{tl}(CC,S)|}
\end{equation}

\begin{lemma} \label{weightcc}
Let $S$ be an assignment to the prefix set $VR$ of variables.\\
Then $Pr({\bf ES}(CC)|{\bf EC}(S)) \leq weight(CC,S)$.
\end{lemma}

{\bf Proof.}
We use induction on the number of variables $Var(CNF(G)) \setminus VR$, i.e. the number of variables
not assigned by $S$. Suppose the number of such variables is $0$. Then none of the clauses of $CC$
is retained w.r.t. $S$. That is, $weight(CC,S)=1$. If all the clauses of $CC$ are satisfied by $S$ then
$Pr({\bf ES}(CC)|{\bf EC}(S))=1$, otherwise it is $0$, hence the lemma holds for the considered case.

Assume now that the number of variables not assigned by $S$ is greater than $0$ and let $X \notin VR$
be a variable such that $VR \cup \{X\}$ is a prefix set (such a variable clearly exists). 
Let $S_1=S \cup \{X\}$ and $S_2=S \cup \{\neg X\}$. 

We say that $X$ is \emph{relevant} to a clause $C$ of $CC$ if 
one of the following is true: (i) $C$ is a tail clause and $X$ occurs in $C$
or (ii) $C=(X_u \vee X_{u,v})$ (that is it is a clause of $MT(CC)$) and
      $X$ is one of $X_u,X_{u,v},X_v$.
\begin{comment}
\begin{itemize}
\item $C$ is a tail clause and $X$ occurs in $C$
\item $C=(X_u \vee X_{u,v})$ (that is it is a clause of $MT(CC)$) and
      $X$ is one of $X_u,X_{u,v},X_v$.
\end{itemize}
\end{comment}
Clearly $X$ can be relevant to at most one clause of $CC$. 
Below we use case analysis to demonstrate that the lemma holds for all types
of relevance of $X$ to $CC$. The cases are organized in a three-level hierarchy.
To make it easier for the reader to go through the rest of the proof we enumerate
the cases in a way that clearly shows that one case is a subcase of another. For example,
case $1.2$ is a subcase of case $1$.
\begin{comment}
By the induction assumption and Lemma \ref{elem2},
$Pr({\bf ES}(CC)|{\bf EC}(S)) \leq weight(CC,S_1)$ if $X=X_{u,v}$ and $\{\neg X_u,\neg X_v\} \subseteq S$ and
$Pr({\bf ES}(CC)|{\bf EC}(S)) \leq weight(CC,S_1)*0.5+weight(CC,S_2)*0.5$.
In the rest of the proof we do case analysis and observe that in each case, the right part of the
appropriate inequality does not exceed $weight(CC)$.
\end{comment}

{\bf Case 1.}
Assume first that $X$ is not relevant to any determining clause of $CC$ w.r.t. $S$
Then $weight(CC,S)=weight(CC,S_1)=weight(CC,S_2)$
because $R_j(CC,S)=R_j(CC,S_1)=R_j(CC,S_2)$ for 
all $j \in \{tl,bc,bf,hc,hf\}$
By the induction assumption and Lemma \ref{elem2},
$Pr({\bf ES}(CC)|{\bf EC}(S)) \leq weight(CC,S_1)$ if $X=X_{u,v}$ and $\{\neg X_u,\neg X_v\} \subseteq S$ and
$Pr({\bf ES}(CC)|{\bf EC}(S)) \leq weight(CC,S_1)*0.5+weight(CC,S_2)*0.5$.
Clearly, the right part of both inequalities does not exceed $weight(CC,S)$. 

{\bf Case 2.}
Assume now that $X$ is relevant to a clause $C \in R_{tl}(CC,S)$.
Then $R_{tl}(CC,S_1)=R_{tl}(CC,S_2)=R_{tl}(CC,S) \setminus \{C\}$ and 
$R_j(CC,S)=R_j(CC,S_1)=R_j(CC,S_2)$ for 
all $j \in \{bc,bf,hc,hf\}$.
It follows that $weight(CC,S_1)=weight(CC,S_2)=2*weight(CC,S)$.

Assume that $C=(X)$. Then $S_2$ falsifies $C$ and hence 
$Pr({\bf ES}(CC)|{\bf EC}(S_2))=0$. Hence, by Lemma \ref{elem2} and the induction assumption,
$Pr({\bf ES}(CC)|{\bf EC}(S)) \leq weight(CC,S_1)*0.5=weight(CC,S)$.
If $C=(\neg X)$, the reasoning is symmetric.

{\bf Case 3.}
It remains to assume that $X$ is relevant to a clause $C=(X_u, \vee X_{u,v})$ of $MT(CC)$.

{\bf Case 3.1.}
Assume first that $X=X_{u,v}$. As $XV \cup \{X\}$ is a prefix set, both $X_u$ and $X_v$ occur in $S$.
Moreover, as $C$ is retained and non-resolved, $\{\neg X_u,X_v\} \subseteq S$. It follows
that $C \in R_{hf}(CC,S)$. Clearly, $R_{hf}(CC,S_1)=R_{hf}(CC,S_2)=R_{hf}(CC,S) \setminus \{C\}$ and 
$R_j(CC,S)=R_j(CC,S_1)=R_j(CC,S_2)$ for 
all $j \in \{tl,bc,bf,hc\}$
It follows that $weight(CC,S_1)=weight(CC,S_2)=2*weight(CC,S)$.
As $S_2$ falsifies $C$, $Pr({\bf ES}(CC)|{\bf EC}(S_2))=0$. Hence, by Lemma \ref{elem2} and the induction assumption,
$Pr({\bf ES}(CC)|{\bf EC}(S)) \leq weight(CC,S_1)*0.5=weight(CC)$ thus confirming the lemma for the considered case.

{\bf Case 3.2.}
Assume now that $X=X_u$. 
%Then $X_{u,v}$ does not occur in $XV$ indeed by definition of a prefix set,
%$X_{u,v} \in XV$ implies $X_u \in XV$ in contradiction to our assumption that $X \notin X
That is, $X_u \notin VR$ hence $\neg X_u \notin S$ and hence, in turn $C$ is a binary clause,
that is $C \in R_{bc}(CC,S) \cup R_{bf}(CC,S)$.

{\bf Case 3.2.1.}
Assume first that $C \in R_{bc}(CC,S)$. Then $C$ is satisfied by $S_1$ and hence 
$R_{bc}(CC,S_1)=R_{bc}(CC,S) \setminus \{C\}$ and 
$R_j(CC,S)=R_j(CC,S_1)$ for 
all $j \in \{tl,bf,hc,hf\}$.
It follows that $weight(CC,S_1)=8/7*weight(CC,S)$.

Note that $C$ is not falsified by $S_2$ because $X_{u,v}$ does not occur in $VR$ 
(indeed, as $VR$ is a prefix set, $X_{u,v} \in VR$ implies $X_u \in VR$ in contradiction to our assumption that $X \notin VR$).
It follows that $C \in R_{hc}(CC,S_2)$ ($X_v \notin S$ implies $X_v \notin S_2$). 
That is, $R_{bc}(CC,S_2)=R_{bc}(CC,S) \setminus \{C\}$ and
$R_{hc}(CC,S_2)=R_{hc}(CC,S) \cup \{C\}$ and
$R_j(CC,S)=R_j(CC,S_2)$ for 
all $j \in \{tl,bf,hf\}$.
Consequently, $weight(CC,S_2)=weight(CC,S)*8/7*3/4=weight(CC,S)*6/7$.
Combining Lemma \ref{elem2} and the induction assumption, we obtain
$Pr({\bf ES}(CC)|{\bf EC}(S)) \leq weight(CC,S_1)*0.5+weight(CC,S_2)*0.5=
weight(CC,S)*8/7*1/2+weight(CC,S)*6/7*1/2=weight(CC,S)$
as required.

{\bf Case 3.2.2.}
Assume now that $C \in R_{bf}(CC,S)$. Applying the same line of reasoning as in the previous
paragraph, we first observe that $R_{bf}(CC,S_1)=R_{bf}(CC,S) \setminus \{C\}$
and $R_j(CC,S)=R_j(CC,S_1)$ for 
all $j \in \{tl,bc,hc,hf\}$.
It follows that $weight(CC,S_1)=4/3*weight(CC,S)$.
Then we observe that $R_{bf}(CC,S_2)=R_{bf}(CC,S) \setminus \{C\}$ and
$R_{hf}(CC,S_2)=R_{hf}(CC,S) \cup \{C\}$ and
$R_j(CC,S)=R_j(CC,S_2)$ for 
all $j \in \{tl,bc,hc\}$.
Consequently, $weight(CC,S_2)=weight(CC,S)*4/3*1/2=weight(CC,S)*2/3$.
Combining Lemma \ref{elem2} and the induction assumption, we obtain
$Pr({\bf ES}(CC)|{\bf EC}(S)) \leq weight(CC,S_1)*0.5+weight(CC,S_2)*0.5=
weight(CC,S)*4/3*1/2+weight(CC,S)*2/3*1/2=weight(CC,S)$
as required.

{\bf Case 3.3.}
It remains to consider the case where $X=X_v$.
As $X_v \notin VR$, $C$ is constrained w.r.t. $S$.

{\bf Case 3.3.1.}
Assume first that $C \in R_{bc}(CC,S)$. 
Then $C$ is binary free w.r.t. $S_1$. That is
$R_{bc}(CC,S_1)=R_{bc}(CC,S) \setminus \{C\}$
and $R_{bf}(CC,S_1)=R_{bf}(CC,S) \cup \{C\}$
and for all $j \in \{tl,hc,hf\}$, $R_j(CC,S_1)=R_j(CC,S)$. 
Thus $weight(CC,S_1)=wegith(CC,S)*6/7$.
Next, $C$ is resolved w.r.t. $S_2$.
That is, $R_{bc}(CC,S_2)=R_{bc}(CC,S) \setminus \{C\}$
and for all $j \in \{tl,bf,hc,hf\}$, $R_j(CC,S_2)=R_j(CC,S)$. 
Thus $weight(CC,S_2)=wegith(CC,S)*8/7$.
Combining Lemma \ref{elem2} and the induction assumption, we obtain
$Pr({\bf ES}(CC)|{\bf EC}(S)) \leq weight(CC,S_1)*0.5+weight(CC,S_2)*0.5=
weight(CC,S)*6/7*1/2+weight(CC,S)*8/7*1/2=weight(CC,S)$
as required.

{\bf Case 3.3.2.}
Finally, assume that $C \in R_{hc}(CC,S)$. 
Applying the line of reasoning as in the previous paragraph,
we observe that $R_{hc}(CC,S_1)=R_{hc}(CC,S) \setminus \{C\}$
and $R_{hf}(CC,S_1)=R_{hf}(CC,S) \cup \{C\}$ and 
for all $j \in \{tl,bc,bf\}$, $R_j(CC,S_1)=R_j(CC,S)$.
Hence  $weight(CC,S_1)=wegith(CC,S)*2/3$.
Next, $C$ is resolved w.r.t. $S_2$.
That is, $R_{hc}(CC,S_2)=R_{hc}(CC,S) \setminus \{C\}$
and for all $j \in \{tl,bf,bc,hf\}$, $R_j(CC,S_2)=R_j(CC,S)$. 
Thus $weight(CC,S_2)=wegith(CC,S)*4/3$.
Combining Lemma \ref{elem2} and the induction assumption, we obtain
$Pr({\bf ES}(CC)|{\bf EC}(S)) \leq weight(CC,S_1)*0.5+weight(CC,S_2)*0.5=
weight(CC,S)*2/3*1/2+weight(CC,S)*=4/3*1/2=weight(CC,S)$
as required.

Observe that the case analysis is complete at this point. Indeed,
a variable $X$ can only be irrelevant to any clause of $DET(CC,S)$
or relevant to a tail clause of $DET(CC,S)$ or relevant to a matching
clause of $DET(CC,S)$. For each of these cases, we have provided exhaustive
classification of subcases and shown that the lemma holds for each of them.
Hence the proof is complete.
$\blacksquare$

Let $M$ be a matching. A \emph{decision tree with a tail} w.r.t. $M$ is a pair
$(DT,TL)$ where $DT$ is a decision tree w.r.t. $M$ and $TL$ is a {\sc cnf} (set of clauses) consisting
of tail clauses (of form $(X_u)$ or $(\neg X_u)$) and the variables occurring in $TL$
do not occur in $DT$ (that is, the variables of $TL$ do not occur as labels of non-leaf nodes of $DT$
nor do they occur in the {\sc cnf}s labelling the leaves of $DST$).
Denote by ${\bf ES}(DT,TL)$ the event consisting of assignment satisfying both $DT$ and $TL$.

\begin{lemma} \label{weighttree}
$Pr({\bf ES}(DT,TL)) \leq (1/2)^{|TL|}*(7/8)^{|M|}$.
\end{lemma}

{\bf Proof.}
By induction on the number of nodes of $DT$. 
Assume that $DT$ has only one node. Then it is associated with $CC$, a matching {\sc cnf} w.r.t. $M$.
Then $CC'=CC \cup TL$ is matching {\sc cnf} with a tail w.r.t. $M$.
Clearly, $Pr({\bf ES}(DT,TL))=Pr({\bf ES}(CC')|EC(\emptyset))$.
Also, it is not hard to see that $R_{tl}(CC',\emptyset)=TL$, $R_{bc}(CC,\emptyset)=CC$
(here $TL$ and $CC$ are treated as sets of clauses) and $R_j(CC,\emptyset)=\emptyset$ for all
other roles $j$. Then the statement of this lemma follows 
by Lemma \ref{weightcc} and, taking into account that $|CC|=|M|$. 

Assume now that $DT$ has more than one node. Let $u$ be the root of $DT$, let $v_1,v_2$
be, respectively, positive and negative children of $u$, and let $DT_1,DT_2$ be the respective
subtrees of $DT$ rooted by $v_1$ and $v_2$. Let $X$ be the variable associated with $u$.
Let $TL_1=TL \cup \{(X)\}$ and $TL_2=TL \cup \{(\neg X)\}$.
By definition of $DT$, $X$ does not occur in $DT_1$, nor in $DT_2$.

Therefore, $(DT_1,TL_1)$ and $(DT_2,TL_2)$ are decision trees with tails
w.r.t. $M$. Note that
${\bf ES}(DT,TL) \subseteq {\bf ES}(DT,TL_1) \cup {\bf ES}(DT,TL_2)$. Indeed, let $S$ be a satisfying
assignment of $CNF(G)$ satisfying $DT$ and $TL$. Assume w.l.o.g. that $X$ is assigned positively.
Then $(u,v_1)$ is the first edge of the path $P$ witnessing that $S$ satisfies $DT$.
Hence, the prefix of $P$ starting at $v_1$ witnesses that $S$ satisfies $DT_1$. 
To see that $S$ satisfies $TL_1$, notice that $TL \subseteq S$ by definition and $X \in S$
because $X$ is the label of $(u,v_1)$.
%Also $X \in S$
%due to the fact that $(u,v_1)$ is an edge of $P$ and all the literal of $TL$ belong to $S$ by definition 
%of a witnessing path. It follows that $S$ satisfies $TL_1$. 
Applying the union bound and then the induction assumption and taking into account that $|TL_1|=|TL_2|=|TL|+1$, 
we get
$Pr({\bf ES}(DT,TL)) \leq Pr(DT,TL_1)+Pr(DT,TL_2) \leq 2*(1/2)^{|TL|+1}*(7/8)^{|M|}=(1/2)^{|TL|}*(7/8)^{|M|}$
as required.
$\blacksquare$

{\bf Proof of Lemma \ref{manytrees}.}
Let $DT$ be a decision tree w.r.t. a matching $M$ of size at least $t$ and let ${\bf ES}(DT)$ be the event 
consisting of the satisfying assignments of $CNF(G)$ satisfying $DT$.
Clearly, ${\bf ES}(DT)={\bf ES}(DT,\emptyset)$ and hence, by Lemma \ref{weighttree},
$Pr({\bf ES}(DT)) \leq (7/8)^t$.

Let ${\bf ES}({\bf A})$ be the event consisting of satisfying assignments of $CNF(G)$ satisfying at least one element of ${\bf A}$. 
Then ${\bf ES}({\bf A}) \leq \bigcup_{DT \in {\bf A}} {\bf ES}(DT)$.
By the union bound and lemma \ref{weighttree},
$Pr({\bf ES}({\bf A}) \leq \sum_{DT \in {\bf A}} {\bf ES}(DT) \leq |{\bf A}|*(7/8)^t$.

Suppose $|{\bf A}|<(8/7)^t$. Then $Pr({\bf ES}({\bf A})<1$. That is, there is a satisfying assignment of $CNF(G)$ that 
does not satisfy any element of ${\bf A}$ in contradiction to the definition of ${\bf A}$.
$\blacksquare$

\section{Proof of Theorem \ref{mainoblivious}} \label{oblsec}
By definition of $k$-{\sc vobp}, there is a sequence $XSV$ of vertex 
variables of $CNF(G)$ where each variable occurs exactly $k$ times 
and such that the sequence of vertex variables occurring on each computational
path of $Z$ is a subsequence of $XSV$. Let $SV$ be the sequence of variables
of $G$ obtained from $XSV$ by replacement of each $X_u$ with $u$.
By definition of $k,c$-fold matching width, there is a matching
$M=\{\{u_1,v_1\}, \dots, \{u_t,v_t\}\}$ of $G$ such that $SV$ has a $c' \leq c$-separation
w.r.t. $U=\{u_1, \dots, u_t\}$ and $V=\{v_1, \dots, v_t\}$. Let $SV_1, \dots SV_{c'}$
be a partition of $SV$ into intervals witnessing the separation. 

Let ${\bf S}$ be the set of all assignments $S$ to the variables of $CNF(G)$
satisfying the following conditions:
(i) $\neg X_{u_i,v_i} \in S$ for each $1 \leq i \leq t$;
(ii) for each $1 \leq i \leq t$, the occurrences of $X_{u_i}$ and
$X_{v_i}$ have distinct signs (that is, the former occurs positively if and only if the latter occurs
negatively);
(iii) the variables besides $\bigcup_{i=1}^t \{X_{u_i},X_{u_i,v_i},X_{v_i}\}$ are assigned 
positively.
\begin{comment}
\begin{itemize}
\item $\neg X_{u_i,v_i} \in S$ for each $1 \leq i \leq t$.
\item For each $1 \leq i \leq t$, the occurrences of $X_{u_i}$ and
$X_{v_i}$ have distinct signs (if the former occurs positively the latter occurs
negatively and if the former occurs negatively the latter occurs positively).
\item The variables besides $\bigcup_{i=1}^t \{X_{u_i},X_{u_i,v_i},X_{v_i}\}$ are assigned 
positively. 
\end{itemize}
\end{comment}
It is not hard to see that each $S \in {\bf S}$ is a satisfying assignment of 
$CNF(G)$. Indeed, for each clause $(X_{u} \vee X_{u,v} \vee X_{v})$,
either $X_{u,v} \in S$ (if $\{u,v\} \neq \{u_i,v_i\}$ for all $1 \leq i \leq t$)
or one of $X_{u},X_{v}$ belongs to $S$ otherwise (due them being assigned oppositely).

It follows that for each $S \in {\bf S}$, we can pick a computational path $P^S$
such that $A(P^S) \subseteq S$. It is not hard to see that there is a partition
$P^S_1, \dots, P^S_{c'}$ into subpaths such that each $SV(P^S_i)$ is a \emph{subsequence} of $SV_i$
(note that we cannot put equality here because a $k$-{\sc vobp} does not have to obey the
uniformity condition). Let $XV^S=(x_1, \dots, x_{c'_1})$ be the sequence of respective ends
of $P^S_1, \dots, P^S_{c'-1}$. 

\begin{lemma} \label{xvdiff}
Let $S_1,S_2$ be two distinct elements of ${\bf S}$. Then $XV^{S_1} \neq XV^{S_2}$.
\end{lemma}

{\bf Proof.}
Assume the opposite and let $S_1,S_2$ be two distinct elements of ${\bf S}$
such that $XV^{S_1}=XV^{S_2}=(x_1, \dots, x_{c'_1})$.
Note that for some $u_i$, $S_1$ and $S_2$ have opposite occurrences of 
$X_{u_i}$. Indeed, otherwise, the occurrences of all $X_{v_i}$ (determined
by $X_{u_i}$) are the same and hence $S_1=S_2$. We assume w.l.o.g.
that there is $u_i$ such that $X_{u_i}$ occurs negatively in $S_1$ and positively in
$S_2$.

Let $P^{S_1}_1, \dots, P^{S_1}_{c'}$ and $P^{S_2}_1, \dots, P^{S_2}_{c'}$
be respective partitions of $P^{S_1}$ and $P^{S_2}$ w.r.t. $x_1, \dots, x_{c'_1}$. 
Let $P^*$ be a root-leaf path of $Z$ passing through $x_1, \dots, x_{c'_1}$
with partition $P^*_1, \dots, P^*_{c'_1}$ into subpaths w.r.t. $x_1, \dots, x_{c'_1}$
such that $P^*_i=P^{S_1}_i$ whenever $i$ is odd and $P^*_i=P^{S_2}_i$ whenever $i$ is 
even. 

Observe that $P^*$ is a computational path. Indeed, all the variables
outsides $XU=\{X_u| u \in U\} \cup XV=\{X_v|v \in V\}$ have the same occurrence
in both $P^{S_1}$ and $P^{S_2}$. If we assume that $X_u \in XU$ has two opposite
occurrences on $P^*$ then such occurrences must happen on some
$P^*_i$ and $P^*_j$ such that $i$ is odd and $j$ is even (otherwise, both these occurrences
happen either on $P^{S_1}$ or on $P^{S_2}$ in contradiction to their consistency).
Notice however that $SV(P^*_j)=SV(P^{S_2}_j)$ is a subsequence of $SV_j$ where $X_u$ does not
occur by definition and hence $X_u$ cannot occur on $P^*_j$, a contradiction. 
The reasoning regarding $X_v \in XV$ is symmetric.

It follows that $P^*$ is a computational path and hence $A(P^*)$ satisfies all the clauses of $CNF(G)$. 
We derive a contradiction by showing that $A(P^*)$ does not satisfy 
$(X_{u_i} \vee X_{u_i,v_i} \vee X_{v_i})$. Indeed, as both $S_1$ and $S_2$ contain $\neg X_{u_i,v_i}$
neither $A(P^{S_1})$ nor $A(P^{S_2})$ contain $X_{u_i,v_i}$ and hence, clearly,
$X_{u_i,v_i} \notin A(P^*)$. Furthermore, $X_{u_i}$ does not belong to $A(P^*_i)$ for an odd $i$
because in this case $A(P^*_i) \subseteq A(P^{S_1}) \subseteq S_1$ and $S_1$ contains $\neg X_{u_i}$.
If $i$ is even $X_{u_i} \notin A(P^*_i)$ simply because, as verified in the previous paragraph,
$X_{u_i}$ does not occur at all on $P^*_i$ for an even $i$. Thus we have shown that $X_{u_i} \notin A(P^*)$.
It can be verified symmetrically that $X_{v_i} \notin A(P^*)$. 
$\blacksquare$

{\bf Proof of Theorem \ref{mainoblivious}}
Let ${\bf XV}=\{XV^S|S \in {\bf S}\}$. It follows from Lemma \ref{xvdiff} that 
$|{\bf XV}| \geq |{\bf S}|$. Observe that $|{\bf S}| \geq 2^t$. 
Indeed, there are $2^t$ distinct assignments to variables $X_{u_1}, \dots, X_{u_t}$.
It is not hard to see that each of these assignments can be extended to an assignment
$S \in {\bf S}$ and that any two assignments obtained this way are distinct, just
because their restrictions to $\{X_{u_1},\dots, X_{u_t}\}$ are distinct. 

It follows that $|{\bf XV}| \geq 2^t$. The rest of the reasoning is analogous 
to the last paragraph of the proof of Theorem \ref{mainupgraded}. 
$\blacksquare$
%\bibliographystyle{plain}
%\bibliography{KnowComp}

\begin{thebibliography}{10}

\bibitem{Ajtai99}
Mikl{\'{o}}s Ajtai.
\newblock A non-linear time lower bound for boolean branching programs.
\newblock In {\em FOCS}, pages 60--70, 1999.

\bibitem{AMaass}
Noga Alon and Wolfgang Maass.
\newblock Meanders and their applications in lower bounds arguments.
\newblock {\em J. Comput. Syst. Sci.}, 37(2):118--129, 1988.

\bibitem{readktimes}
Allan Borodin, Alexander~A. Razborov, and Roman Smolensky.
\newblock On lower bounds for read-k-times branching programs.
\newblock {\em Computational Complexity}, 3:1--18, 1993.

\bibitem{MonComp}
Michelangelo Grigni and Michael Sipser.
\newblock Monotone complexity.
\newblock In {\em Poceedings of the London Mathematical Society Symposium on
  Boolean Function Complexity}, pages 57--75, 1992.

\bibitem{Jukna4clique}
Stasys Jukna and Georg Schnitger.
\newblock Triangle-freeness is hard to detect.
\newblock {\em Combinatorics, Probability {\&} Computing}, 11(6):549--569,
  2002.

\bibitem{RazFCT}
Alexander~A. Razborov.
\newblock Lower bounds for deterministic and nondeterministic branching
  programs.
\newblock In {\em Fundamentals of Computation Theory, 8th International
  Symposium, ({FCT})}, pages 47--60, 1991.

\bibitem{RazgonAlgo}
Igor Razgon.
\newblock On the read-once property of branching programs and cnfs of bounded
  treewidth.
\newblock {\em Algorithmica, to appear}.

\bibitem{RazgonIPEC14}
Igor Razgon.
\newblock No small nondeterministic read-once branching programs for cnfs of
  bounded treewidth.
\newblock In {\em IPEC}, pages 319--331, 2014.

\bibitem{RazgonKR}
Igor Razgon.
\newblock On {O}{B}{D}{D}s for {C}{N}{F}s of bounded treewidth.
\newblock In {\em Principles of Knowledge Representation and Reasoning({KR})},
  2014.

\bibitem{cobdd}
Igor Razgon.
\newblock On oblivious branching programs with bounded repetition that cannot
  efficiently compute {C}{N}{F}s of bounded treewidth.
\newblock {\em CoRR}, abs/1510.02951, 2015.

\bibitem{Sauerhoff03}
Martin Sauerhoff.
\newblock Randomness versus nondeterminism for read-once and read- k branching
  programs.
\newblock In {\em {STACS} 2003, 20th Annual Symposium on Theoretical Aspects of
  Computer Science, Berlin, Germany, February 27 - March 1, 2003, Proceedings},
  pages 307--318, 2003.

\end{thebibliography}

\appendix
\section{Proof of Lemma \ref{unigood}}
%Transformation to a uniform k-VEMBP
%Let $Z$ be a $k$-{\sc vembp} computing $CNF(G)$, let $v$ be a node of $Z$, and let $X$ 
%be a vertex variable. De
Let $Z$ be a $k$-{\sc vembp}. The \emph{in-degree} $d^{+}(v)$ of a node $v$ is a number of in-neighbours (this is essential
point because of the possibility of multiple edges). 
Let us call $Z$ \emph{clean} if the following conditions are true.
\begin{itemize}
\item All the incoming edges of nodes $v$ with $d^{+}(v)>1$ are unlabelled.
\item For each pair of nodes $u,v$, all the $(u,v)$-edges are either unlabelled or labelled with literals
of the same variable.
\end{itemize}
We assume that $Z$ is clean.
This assumption does not restrict generality because a $k$-{\sc vembp} can be 
transformed into a clean one having at most twice more edges than the original $k$-{\sc vembp}
and computing the same function.
Indeed, let $v$ be a node with in-degree greater than $1$ and let $(u,v)$ be an edge labelled with 
a literal $x$. Subdivide $(u,v)$ and let $u,w,v$ be the path that replaced $(u,v)$. Then label $(u,w)$ with $x$.
Clearly, as a result we get a $k$-{\sc vembp} implementing the same function as the original one. 
Notice that $w$ has in-degree $1$, that is the number of edges violating the assumption has decreased by $1$.
Thus, one can inductively argue that in case of $q$ `violating' edges, there is a transformation to a $k$-{\sc vembp}
satisfying the above assumption that creates at most $q$ additional edges. 
As for multiple edges, suppose there are multiple edges $(u,v)$ violating the second condition of the cleanness.
Then subdivide as above all the $(u,v)$ edges labelled with a literal.
As a result we get a clean $k$-{\sc vembp} in which each edge has been subdivided at most once.

Suppose $Z$ computes $CNF(G)$ and let $X$ be a vertex variable of $CNF(G)$.
For a node $v$ of $Z$, denote by $fr_Z(v,X)$ the largest number of occurrences of $X$
on a path from the root to $v$. 
We call the edges $(u,v)$ of $Z$ such that $d^{+}(v)>1$ \emph{relevant}.
Denote the set of relevant edges of $Z$ by $RL_Z$.
A relevant edge is \emph{irregular} w.r.t. $X$ if $fr_Z(v,X)-fr_Z(u,X)>0$. 

Let $(u,v)$ be an edge such that $fr_Z(v,X)-fr_Z(u,X)=q>0$.
Then transform $Z$ as follows. 

\begin{enumerate}
\item Remove the edge $(u,v)$.
\item Introduce new vertices $u_1, \dots, u_q$; we will refer to $u$ as $u_0$ for
the sake of convenience.
\item For each $1 \leq i \leq q$, introduce two edges $(u_{i-1},u_i)$ and label
them them $X$ and $\neg X$, respectively.
\item Introduce an unlabelled edge $(u_q,v)$.
\end{enumerate}

Let $Z'$ be the graph obtained as a result of the above transformation.
The following properties can be observed by a direct inspection. 
%The following facts are easy to observe by construction.
\begin{observation} \label{easyfacts}
\begin{enumerate}
\item $fr_Z(v,X)=fr_{Z'}(X)$.
\item The edge $(u_q,v)$ is regular w.r.t. $X$ in $Z'$.
\item $Z'$ is clean. 
\item $Z$ and $Z'$ have the same set of nodes with in-degree greater than one.
\item For each vertex variable $Y \neq X$, and node $w$ of 
$Z$, $fr_Z(w,Y)=fr_{Z'}(w,Y)$. Moreover, $fr_Z(u,Y)=fr_Z(u_q,Y)$. 
\item $RL_{Z'}=RL_Z \setminus \{(u,v)\} \cup \{(u_q,v)\}$
\end{enumerate}
\end{observation}

\begin{lemma} \label{remainskvembp}
$Z'$ is a $k$-{\sc vembp} computing the same function as $Z$.
\end{lemma}

{\bf Proof.}
Clearly, the transformation from $Z$ to $Z'$ preserves monotonicity of 
edge variables of $CNF(G)$. 
In light of statement 4 of Observation \ref{easyfacts}, it is sufficient to establish
that there are no more than $k$ occurrences of $X$
on each root-leaf path $P$ of $Z'$
that is not a path of $Z$.
By construction, such a path $P$ includes $u$ and $v$
and the subpath $P_{u,v}$ starting at $u$ and ending at $v$ goes
through $u_1, \dots, u_q$ as defined above.
Let $P_u$ be the prefix of $P$ ending at $u$,
and $P_v$ be the suffix of $P$ beginning at $v$.
By definition of the $fr$ function, the number of occurrences of 
$X$ on $P_u$ is at most $fr_Z(v)-q$. Furthermore, the number of occurrences
of $X$ on $P_v$ is at most $k-fr_Z(v)$. Indeed, otherwise, let $P'$ be a root-$v$
path witnessing $fr_Z(v)$. Then, appending $P_v$ to the end of $P'$ we obtain
a root-leaf path of $Z$ with more than $k$ occurrences of $X$, a contradiction.
Thus, the number of occurrences of $X$ on $P$ is at most $fr_Z(v,X)-q$ on
$P_u$ plus $q$ occurrences of $P_{u,v}$ plus at most $k-fr_Z(v,X)$ occurrences on $P_v$.
Clearly, on $P$, there are at most $k$ occurrences of $X$ in total. 

Let $S$ be a satisfying assignment of the function computed by $Z$
and let $P$ be a computational path of $Z$ with $A(P) \subseteq S$.
If $P$ does not include $(u,v)$ then $P$ is a computational path of $Z'$.
Otherwise, let $P_u$ and $P_v$ be as in the previous paragraph 
and let $P'$ be a $u-v$ path with $u_1, \dots, u_q$ being the intermediate vertices
and the in-edge for each $u_i$ is the one labelled with the literal of $X$ that
belongs to $S$ (by construction, such a selection is possible) and, as a result $A(P') \subseteq S$.
Taking into account that $A(P_u) \cup A(P_v) \subseteq A(P) \subseteq S$,
we conclude that $A(P_u+P'+P_v) \subseteq S$.
That is, in any case there is a computational path of $Z$ whose set of literals is a subset of $S$ and hence 
$S$ is a satisfying assignment of the function computed by $Z'$.

Conversely, let $S$ be a satisfying assignment of the function computed by $Z'$.
Let $P$ be a computational path of $Z'$ such that $A(P) \subseteq S$. If $P$ is not
a path of $Z$ then, by construction, $P$ includes both $u$ and $v$ and a path of
$Z$ can be obtained by replacement of the subpath of $P$ between $u$ and $v$ by an
edge $(u,v)$. Clearly, the set of literals of this resulting path is a subset of $A(P)$,
hence $S$ is a satisfying assignment of the function computed by $Z$.
$\blacksquare$

Denote by $RG_Z(Y)$, $IR_Z(Y)$ the respective sets of regular and 
irregular edges of $Z$ w.r.t. a vertex variable $Y$.

\begin{lemma} \label{moreregular}
$|IR_{Z'}(X)|<|IR_Z(X)|$ and for each $Y \neq X$, $|IR_{Z'}(Y)| \leq |IR_Z(Y)|$.
\end{lemma}

{\bf Proof.}
Statement 6 of Observation \ref{easyfacts} lets us define a bijection from
the relevant edges of $Z$ to the relevant edges of $Z'$ so that $(u,v)$ corresponds to $(u_q,v)$
and each other edge corresponds to itself. Observe that
for each $Y \neq X$ each regular edge of $Z$ w.r.t. $Y$ corresponds to a regular edge of
$Z'$ w.r.t. $Y$. Indeed, assume that $(u,v)$ is regular w.r.t. $Y$.
Then, by the Statement 5 of Observation \ref{easyfacts},
$fr_{Z'}(v,Y)-fr_{Z'}(u_q,Y)=fr_Z(v,Y)-fr_Z(u,Y)=0$. That is, the corresponding edge $(u_q,v)$ of $Z'$
is also regular w.r.t. $Y$. For other regular edges, the argumentation is similar.
It follows that $|RG_{Z'}(Y)| \geq |RG_Z(Y)|$. Then, since $|RL_Z|=|RL_{Z'}|$,
it follows that
$|IR_{Z'}(Y)|=|RL_{Z'} \setminus RG_{Z'}(Y)| \leq |RL_{Z} \setminus RG_Z(Y)|=|IR_Z(Y)|$,
proving the second statement.

For the first statement, we need an additional claim.
\begin{claim}
For each node $w$ of $Z$, $fr_Z(w,X)=fr_{Z'}(w,X)$. 
\end{claim}

{\bf Proof.}
By the first statement of Observation \ref{easyfacts}, the claim holds for $v$.
For $w \neq v$, it is not hard to see that 
$fr_Z(w,X) \leq fr_{Z'}(w,X)$. We are going to establish that 
$fr_Z(w,X) \geq fr_{Z'}(w,X)$. Let $P$ be a root-$w$ path of $Z'$ with the largest number
of occurrences of $X$. If $P$ is also a path of $Z$ then $fr_Z(w,X) \geq fr_{Z'}(w,X)$
follows immediately. Otherwise, $P$ passes through $v$. Let $P_v$ be the prefix of $P$
ending at $v$. Let $P^*$ be a root-$v$ path of $Z$ witnessing $fr_Z(v,X)$. 
As $fr_Z(v,X)=fr_Z'(v,X)$, the number of occurrences of $X$ on $P^*$ is not smaller than
on $P_v$. Due to the irregularity of $(u,v)$, it is not an edge of $P^*$ and hence $P^*$ is 
also a path of $Z'$. It follows that, transforming $P$ by replacing $P_v$ with $P^*$,
we obtain a new root-$w$ path $P'$ of $Z'$ with a number of occurrences of $X$ at least
$fr_{Z'}(w,X)$. As $P'$ is also a path of $Z$, it follows that 
$fr_Z(w,X) \geq fr_{Z'}(w,X)$. $\square$

It follows that the above bijection from $RL_Z$ to $RL_{Z'}$ maps each regular
edge of $Z$ w.r.t. $X$ to a regular edge of $Z'$ w.r.t. $X$. Indeed, let $(u',v')$ be a regular edge of
$Z$ w.r.t. $X$. By assumption about $(u,v)$, $(u',v') \neq (u,v)$ and hence the bijection
maps $(u',v')$ to itself. 
It follows from the claim that
$fr_{Z'}(v',X)-fr_{Z'}(u',X)=fr_Z(v',Y)-fr_Z(u',Y)=0$ and hence $(u',v')$ is regular w.r.t. $X$ in 
$Z'$. In addition, by Observation \ref{easyfacts}, $(u,v)$, an irregular edge w.r.t. $X$ in $Z$ 
corresponds to a regular edge $(u_q,v)$ w.r.t. $X$ in $Z'$. It follows that
$|RG_Z(X)|<|RG_{Z'}(X)|$ and hence the first statement follows by a calculation similar to the one
we used for the second statement. $\blacksquare$

{\bf Proof of Lemma \ref{unigood}.}
Suppose that $Z$ is clean. 
Denote $\sum_{X \in Vvar(G)} |IR_Z(X)|$ by $ir_Z$.
We prove, by induction on $ir_Z$, 
that by adding at most $ir_Z*k$ new vertices,
$Z$ can be transformed into a $k$-{\sc vembp} $Z^*$
where all the relevant edges are regular w.r.t. all the variables.

If $ir_Z=0$ then no transformation is needed.
Otherwise, pick an edge $(u,v)$ irregular w.r.t. a variable $X$
and apply transformation as above. Let $Z'$ be the resulting
graph. Then, by Lemma \ref{moreregular},
$ir_{Z'}<ir_Z$. By Observation \ref{easyfacts}, $Z'$ is clean and 
by Lemma \ref{remainskvembp}, $Z'$ is a $k$-{\sc vembp}
computing $CNF(G)$. By the induction assumption, $Z'$ can be transformed
to $Z^*$ as above by adding at most $ir_{Z'}*k$ vertices.
Hence, the transformation from $Z$ requires at most
$(ir_{Z'}+1)*k \leq ir_Z*k$ new vertices as required. 

Observe that for each node $v$ of $Z^*$ and each vertex variable $X$,
all root-$v$ paths of $Z^*$ carry the $fr_{Z^*}(v,X)$ occurrences of 
$X$. Indeed, assume that this is not so and let $P$ be the shortest
path violating this statement w.r.t. a variable $X$. 
Let $w$ be the last node of $P$. Then the number of occurrences of $X$
on $P$ is smaller than $fr_{Z^*}(w,X)$.

Assume first that $w$ has in-degree $1$ and let $u$ be the only 
in-neighbour. Then, it follows from the second condition of a clean $k$-{\sc vembp},
the number of occurrences of $X$ on $P_u$, the prefix of $P$ ending at $u$
is smaller than $fr_Z(u,X)$ in contradiction to the minimality of $P$.
Assume now that the in-degree of $w$ is larger than $1$ and let $P'$ be a root-$w$ path
witnessing $fr_{Z^*}(w,X)$. If $P$ and $P'$ have the same penultimate vertex
then we get the same contradiction with the minimality of $P$ as in the previous case.
Otherwise, we conclude that the last edge of $P$ is irregular w.r.t. $X$,
in contradiction to $ir_{Z^*}=0$. 

Let $lf$ be the leaf of $Z^*$. If for each variable $X$, $fr_{Z^*}(lf,X)=k$
then we are done. Otherwise, subdividing in-coming edges of $lf$
as described in the above transformation, we can add the required number of 
occurrences of each variable $X$ with $fr_{Z^*}(lf,X)<k$.

Clearly, during the above transformation, each edge of $Z$ is subdivided at most $O(nk)$ times.
Hence, the total number of nodes in the resulting $k$-{\sc vembp}
is $O(mnk)+|Z|$ where $m$ is the number of edges of $Z$. 
Now, what is an upper bound on $m$ in terms of $n$ and $Z$? 
The number of multiple edges between any
two particular vertices $u$ and $v$ of $Z$ can be assumed $O(n)$
because if there are two multiple edges unlabelled or two labelled
with the same literal then one of them can be safely removed. 
Then $m=O(nZ^2)$. It follows that for any $k$-{\sc vembp}
there is a uniform $k$-{\sc vembp} with $O(n^2|Z|^2k)$ nodes as required. 
$\blacksquare$

\section{Proofs omitted from Section 4.2.}
\begin{comment}
\begin{observation} \label{elem1}
Let $S$ be a set of literals such that $Var(S)$ is a prefix set.
Let $X \notin Var(S)$ be a variable such that $Var(S) \cup \{X\}$
is still a prefix set. 
Let $S_1=S \cup \{X\}$ and $S_2=S \cup \{\neg X\}$. 
Suppose that $X=X_{u,v}$ and both $X_u$ and $X_v$ occur negatively in $S$.
Then $Pr({\bf EC}(S_1)|{\bf EC}(S))=1$ and $Pr({\bf EC}(S_2)|{\bf EC}(S))=0$.
Otherwise, both $Pr({\bf EC}(S_1)|{\bf EC}(S))=0.5$ and
$Pr({\bf EC}(S_2)|{\bf EC}(S))=0.5$.
\end{observation}
\end{comment}

{\bf Proof of Observation \ref{elem1}}
First of all, we need an auxiliary claim.

\begin{claim}
Let $S$ be a set of literals such that $Var(S)$ is a prefix set.
Let $Y \in S$ be a literal of a variable $X$.
Let us define the individual probability $pr_S(Y)$ of $Y$ (w.r.t. $S$) as follows.
\begin{itemize}
\item If $X$ is a vertex variable then $pr(Y)=0.5$.
\item If $X=X_{u,v}$ and either $X_u \in S$ or $X_v \in S$ then $pr(Y)=0.5$.
\item If $X=X_{u,v}$ and neither of $X_u$ and $X_v$ belongs to $S$
(that is, both of the negations belong to $S$, by definition of a prefix set)
then $pr(Y)=1$ if $Y=X$ and $pr(Y)=0$ if $Y=\neg X$.
\end{itemize}
Denote $\prod_{Y \in S} pr_S(Y)$ by $pr(S)$.
Then $Pr(EC(S))=pr(S)$.
\end{claim}

{\bf Proof.}
By induction on the number of variables not assigned by $S$. 
If the number of such variables is $0$ then $S$ assigns all the variables of $CNF(G)$.
In this case the claim follows by definition of the probability space.
Assume now that $S$ does not assign all the variables of $CNF(G)$ and let 
$X \in Var(CNF(G)) \setminus Var(S)$
such that $Var(S) \cup \{X\}$ is a prefix set
(it is not hard to see that such a variable always
exists). Then, clearly ${\bf EC}(S)$ is the disjoint union of ${\bf EC}(S \cup \{X\})$
and ${\bf EC}(S \cup \{\neg X\})$. That is, 
$Pr({\bf EC}(S))=Pr({\bf EC}(S \cup \{X\}))+Pr({\bf EC}(S \cup \{\neg X\}))$.
By the induction assumption,
$Pr({\bf EC}(S \cup \{X\}))=pr(S \cup \{X\})$ and 
$Pr({\bf EC}(S \cup \{\neg X\}))=pr(S \cup \{\neg X\})$.

By definition, $pr(S \cup \{X\})=(\prod_{Y \in S} pr_{S \cup \{X\}} Y)*
pr_{S \cup \{X\}} X$.
Notice that for each $Y \in S$, $pr_{S \cup \{X\}} Y=pr_S Y$.
This is certainly true if $Y$ is a literal of a vertex variable: the individual 
probability of $Y$ is always $0.5$. If $Y$ is a literal of an edge variable $X_{u,v}$
then, due to being $Var(S)$ a prefix set, literal of $X_u$ and $X_v$ belong to $S$
and hence \emph{the very same literals} belong to $S \cup X$, confirming the observation.
Thus we can write, 
$Pr({\bf EC}(S \cup \{X\}))=(\prod_{Y \in S} pr_{S} Y)*pr_{S \cup \{X\}} X$
and, analogously,  
$Pr({\bf EC}(S \cup \{\neg X\}))=(\prod_{Y \in S} pr_{S} Y)*pr_{S \cup \{\neg X\}} X$
From the definition of individual probabilities, it is clear that 
$pr_{S \cup \{X\}} X+pr_{S \cup \{\neg X\}} X=1$. Hence, adding up 
$Pr({\bf EC}(S \cup \{X\}))$ and $Pr({\bf EC}(S \cup \{\neg X\}))$ results in $pr(S)$,
as required. 
$\square$ 

Observe that since ${\bf EC}(S) \subset {\bf EC}(S_1)$ and
${\bf EC}(S) \subset {\bf EC}(S_2)$,
$Pr({\bf EC}(S_1) \cup {\bf EC}(S))=Pr({\bf EC}(S_1))$
and $Pr({\bf EC}(S_2) \cup {\bf EC}(S))=Pr({\bf EC}(S_2))$.
By definition of conditional probability and the above claim,
$Pr({\bf EC}(S_1)|{\bf EC}(S))=Pr({\bf EC}(S_1))/Pr({\bf EC}(S))=
pr(S_1)/pr(S)$ and, analogously, 
$Pr({\bf EC}(S_2)|{\bf EC}(S))=pr(S_2)/pr(S)$. The observation now immediately
follows from definition of individual probability in the above claim.
$\blacksquare$

{\bf Proof of Lemma \ref{elem2}.}
It is not hard to see that ${\bf EC}(S)$ is the disjoint union of ${\bf EC}(S_1)$ and
${\bf EC}(S_2)$.
Therefore, by the law of full probability,
$Pr({\bf X} \cap {\bf EC}(S))=Pr({\bf X} \cap {\bf EC}(S_1))+Pr({\bf X} \cap {\bf EC}(S_2))$.
Or, in terms of conditional probabilities
\begin{equation} \label{eq1}
 Pr({\bf X}|{\bf EC}(S))*Pr({\bf EC}(S))=
 Pr({\bf X}|{\bf EC}(S_1))*Pr({\bf EC}(S_1))+Pr({\bf X}|{\bf EC}(S_2))*Pr({\bf EC}(S_2))
\end{equation}

Notice that since ${\bf EC}(S_1) \subseteq {\bf EC}(S)$, we can write
\begin{equation} \label{eq2}
Pr({\bf EC}(S_1))=Pr({\bf EC}(S_1) \cap {\bf EC}(S))=Pr({\bf EC}(S_1)|{\bf EC}(S))*Pr({\bf EC}(S))
\end{equation}

Likewise,
\begin{equation} \label{eq3}
Pr({\bf EC}(S_2))=Pr({\bf EC}(S_2)|{\bf EC}(S))*Pr({\bf EC}(S))
\end{equation}

Substituting (\eqref{eq2}) and (\eqref{eq3}) into the right hand side of (\eqref{eq1})
and dividing both sides by $Pr({\bf EC}(S))$ gives us

\begin{equation} \label{eq4}
Pr({\bf X}|{\bf EC}(S))=
Pr({\bf X}|{\bf EC}(S_1))*Pr({\bf EC}(S_1)|{\bf EC}(S))+Pr({\bf X}|{\bf EC}(S_2))*Pr({\bf EC}(S_2)|{\bf EC}(S))
\end{equation}

The Lemma now follows from Observation \ref{elem1} by choosing the appropriate $Pr({\bf EC}(S_1)|{\bf EC}(S))$
and $Pr({\bf EC}(S_2)|{\bf EC}(S))$ for each considered case.
$\blacksquare$
\end{document}